\documentclass[journal]{IEEEtran}
\usepackage{caption}
\usepackage{epsfig}
\usepackage{epstopdf}
\usepackage{hyperref}
\usepackage{changepage}
\usepackage{tabularx}
\usepackage{array}
\usepackage{textcomp} 
\usepackage{pifont}
\usepackage{graphicx}
\usepackage{siunitx}
\usepackage{booktabs}
\usepackage{xstring}
\usepackage[T1]{fontenc}
\usepackage{hyperref}
\usepackage{xcolor} 
\usepackage{amssymb}  
\UseRawInputEncoding  
\usepackage{caption}
\usepackage{multirow} 
\usepackage{makecell} 
\newcommand{\bea}{\begin{eqnarray}}
\newcommand{\eea}{\end{eqnarray}}

\newcommand{\be}{\begin{equation}}
\newcommand{\ee}{\end{equation}}

\usepackage{algorithmic}
\usepackage{algorithm}
\usepackage{mathrsfs}
\usepackage{amsfonts,dsfont}
\usepackage{amssymb,colortbl}
\usepackage{graphicx}
\usepackage{subfigure}
\usepackage{amsmath,booktabs,latexsym,verbatim}

\begin{document}
\title{Defending Against Network Attacks for Secure AI Agent Migration in Vehicular Metaverses}

\author{Xinru Wen, Jinbo Wen, Ming Xiao, Jiawen Kang, Tao Zhang, Xiaohuan Li,\\ Chuanxi Chen, Dusit Niyato, \textit{Fellow, IEEE}
\thanks{ 
        X. Wen, M. Xiao, and J. Kang are with the School of Automation, Guangdong University of Technology, Guangzhou 510006, China (e-mails: 2112304038@mail2.gdut.edu.cn; xiaoming@gdut.edu.cn; kavinkang@gdut.edu.cn).
        J. Wen is with the College of Computer Science and Technology, Nanjing University of Aeronautics and Astronautics, Nanjing 210016, China (e-mail: jinbo1608@163.com).
        T. Zhang is with the School of Cyberspace Science and Technology, Beijing Jiaotong University, Beijing 100044, China (e-mail: taozh@bjtu.edu.cn). 
        X. Li is with the College of Information and Communication, Guilin University of Electronic Technology, Guilin 541004, China (e-mail: lxhguet@guet.edu.cn).
        C. Chen is with the College of Computer and Cyber Security, Fujian Normal University, Fuzhou 350007, China (e-mail: cxchen@fjnu.edu.cn).
        D. Niyato is with the College of Computing and Data Science, Nanyang Technological University, Singapore (e-mail: dniyato@ntu.edu.sg).
        \textit{Corresponding authors: Jiawen Kang, Chuanxi Chen.}
	}}
\maketitle

\begin{abstract}
Vehicular metaverses, blending traditional vehicular networks with metaverse technology, are expected to revolutionize fields such as autonomous driving. As virtual intelligent assistants in vehicular metaverses, Artificial Intelligence (AI) agents powered by large language models can create immersive 3D virtual spaces for passengers to enjoy on-broad vehicular applications and services. To provide users with seamless and engaging virtual interactions, resource-limited vehicles offload AI agents to RoadSide Units (RSUs) with adequate communication and computational capabilities. Due to the mobility of vehicles and the limited coverage of RSUs, AI agents need to migrate from one RSU to another RSU. However, potential network attacks pose significant challenges to ensuring reliable and efficient AI agent migration. In this paper, we first explore specific network attacks including traffic-based attacks (i.e., DDoS attacks) and infrastructure-based attacks (i.e., malicious RSU attacks). Then, we model the AI agent migration process as a Partially Observable Markov Decision Process (POMDP) and apply multi-agent proximal policy optimization algorithms to mitigate DDoS attacks. In addition, we propose a trust assessment mechanism to counter malicious RSU attacks. Numerical results validate that the proposed solutions effectively defend against these network attacks and reduce the total latency of AI agent migration by approximately 43.3\%.
\end{abstract}


\IEEEpeerreviewmaketitle

\section{Introduction}
\label{intro}

Metaverses represent digital environments that incorporate advanced technologies, including Extended Reality (XR), Digital Twins (DTs), Generative Artificial Intelligence (GAI), and Large Language Models (LLMs) \cite{9944868, 10254627}. By using head-mounted displays, users can engage with virtual environments and obtain immersive experiences, achieving seamless integration between physical and virtual spaces \cite{10254627}. With the rapid development of the Internet of Vehicles (IoV) and widespread applications of metaverse technologies, vehicular metaverses are gradually becoming a new paradigm of intelligence and digitalization, which are regarded as comprehensive digital ecosystems that integrate advanced AI, and vehicular networking technologies \cite{10401029}, enabling immersive interactions between vehicles and surrounding virtual infrastructures within simulated environments.
As virtual intelligent assistants powered by LLMs in vehicular metaverses, AI agents cover a variety of functions beyond natural language processing, including decision-making, problem-solving, interacting with external environments, and performing actions \cite{zhang2024}. Specifically, AI agents begin by collecting and processing vehicular information from their surroundings and then generate potential strategies based on resource availability and network conditions. Subsequently, AI agents perform actions and deliver personalized vehicular services to users, such as virtual entertainment and real-time navigation \cite{9973495}.


The construction and continuous updates of AI agents require intensive computational and storage resources. For example, the computational power required to build AI agents similar to GPT-3 may range from tens to thousands of tera floating point operations per second \cite{zhang2024}. To ensure real-time and immersive AI agent services, AI agents can be deployed on RoadSide Units (RSUs), enabling users to interact through voice commands or other interfaces \cite{9880566}. Given the mobility of vehicles and the limited coverage provided by a single RSU, it is essential to execute real-time online migration of AI agents to maintain high-quality AI agent services for users. This process involves dynamically migrating AI agents from one RSU to another one when vehicles move \cite{YuezhongMetaverse}. To enhance the efficiency of AI agent migration, pre-migration strategies play a vital role \cite{9880566}, which involves proactively transferring all or part of AI agent services and their related data to the subsequent RSUs before vehicles enter their coverage area. 

However, ensuring the security of AI agent migration remains a critical challenge in vehicular metaverses. Network attackers may launch Distributed Denial of Service (DDoS) attacks against RSUs to disrupt AI agent migration or deploy malicious RSUs to collect data uploaded by vehicles \cite{10.1007/978-981-16-8664-1_16}. These attacks pose significant threats to the integrity of AI agent migration. For instance, by sending a flood of requests, DDoS attacks overwhelm communication channels between AI agents and their linking RSUs, hindering the timely and secure AI agent migration, which may cause connection timeouts or even complete service disruptions between users and AI agents \cite{jsan12040051}. Moreover, AI agent migration may be compromised by malicious RSU attacks, where RSUs compromised or even controlled by attackers steal sensitive information and manipulate vehicle data, potentially misleading the decision-making processes of vehicles during migration. 

Currently, most research on DDoS defense focuses on traditional vehicular network scenarios \cite{9829332, 10018506, 8293801, 10.1007/s10586-023-04035-5}, while studies addressing DDoS attacks during DT migration in vehicular metaverses remain limited \cite{kang2024}. Although some research explores this emerging field, these studies often assume simplified DDoS attack models and lack comprehensive defense strategies capable of countering diverse types of attacks. Additionally, existing studies do not provide strong defenses against malicious RSUs that can compromise user privacy. To address these challenges, we propose a secure online AI agent migration framework coupled with a trust assessment mechanism, which can effectively mitigate DDoS and malicious RSU attacks while enabling online AI agent migration with minimal latency. Our contributions are summarized as follows:
\begin{itemize}
\item We propose a new online AI agent migration framework in vehicular metaverses aimed at providing highly immersive services for vehicular users. For this framework, we introduce traffic-based attacks (e.g., direct/indirect and hybrid DDoS attacks) and infrastructure-based attacks (e.g., malicious RSU attacks) during AI agent migration, and implement the modeling of attack models.

\item For traffic-based attacks, we consider the challenges of DDoS defense and minimize the total latency of AI agent migration into an optimization problem. Then, we model the proposed optimization problem as a Partially Observable Markov Decision Process (POMDP) and leverage Multi-Agent Proximal Policy Optimization (MAPPO) algorithms to learn the optimal AI agent pre-migration strategy, thus resisting DDoS attacks while minimizing the total latency of AI agent migration. 

\item For infrastructure-based attacks, we develop a trust assessment mechanism to mitigate malicious RSU attacks during AI agent migration. Specifically, we first define the malicious score of RSUs based on anomaly detection and performance during the migration task. We then propose an algorithm to calculate the malicious score of RSUs, update the malicious threshold, and prohibit malicious RSUs from executing AI agent migration.
\end{itemize}

The remainder of this paper is organized as follows: Section \ref{sec:rl} reviews the related work. Section \ref{sec:sy} presents the online AI agent migration framework in vehicular metaverses. In Section \ref{sec:pf}, we formulate the optimization problem for DDoS defense during AI agent migration. Section \ref{sec:sa} provides the details of the proposed MAPPO-based online AI agent pre-migration algorithm. Section \ref{sec:ta} presents the trust assessment mechanism to mitigate malicious RSU attacks. Section \ref{sec:e} discusses numerical results. Finally, Section \ref{sec:c} concludes the paper.

\section{Related Work}
\label{sec:rl}

In this section, we review the works related to vehicular metaverses, twin migration, and network attacks in vehicular networks. Furthermore, we summarize the differences between the related works and our work in Table \ref{comparison}.

\subsection{Vehicular Metaverses}

\begin{table*}
    \setlength{\abovecaptionskip}{0pt}%
    \setlength{\belowcaptionskip}{-1em}%
    \caption{Comparisons between related works and our work}
    \label{comparison}
    \centering
    \setlength{\tabcolsep}{5.0pt} 
    \begin{tabular}{|c|c|c|c|c|c|c|c|}
        \hline
        \rule{0pt}{8pt}
        \begin{tabular}{c} \multirow{4}*{Reference} \end{tabular}
        & \multicolumn{3}{c|}{\centering \textbf{Optimization Objectives}} 
        & \multicolumn{2}{c|}{\centering \textbf{Attack Types}} 
        & \multicolumn{2}{c|}{\centering \textbf{Defense Methods}}\\
        \cline{2-8}
         & \begin{tabular}{c} 
                Latency
          \end{tabular}
        & \begin{tabular}{c} 
                Offloaded\\tasks 
          \end{tabular}
        & \begin{tabular}{c} 
                Safety
          \end{tabular}
          & \begin{tabular}{c} 
                DDoS\\attacks
          \end{tabular}
          & \begin{tabular}{c} 
             Malicious RSU\\attacks
          \end{tabular}
          & \begin{tabular}{c} 
                Deep learning,\\ machine learning,\\reinforcement learning
          \end{tabular}
          & \begin{tabular}{c} 
                 Reputation\\assessment\\mechanism
          \end{tabular}\\
        \hline 
        \cite{9944868}  & \ding{53} & $\checkmark$  & $\checkmark$ & \ding{53} & \ding{53} & \ding{53} & \ding{53}\\
        \hline 
        \cite{9880566}  & \ding{53} & $\checkmark$  & $\checkmark$ & \ding{53} & \ding{53} & \ding{53} &$\checkmark$\\
        \hline 
        \cite{9829332} & \ding{53} & $\checkmark$ & $\checkmark$ & $\checkmark$ & $\checkmark$ & $\checkmark$ &$\checkmark$\\
        \hline 
        \cite{10018506}  & \ding{53} & \ding{53} & $\checkmark$ & $\checkmark$ & \ding{53} & $\checkmark$ &\ding{53}\\
        \hline 
        \cite{8293801}  & \ding{53} & \ding{53}  & $\checkmark$ & $\checkmark$ & \ding{53} & \ding{53} &\ding{53}\\
        \hline 
        \cite{10.1007/s10586-023-04035-5}  & \ding{53} & $\checkmark$ &  $\checkmark$ & \ding{53} & $\checkmark$ & $\checkmark$ &\ding{53}\\
        \hline 
        \cite{kang2024}   & $\checkmark$ & $\checkmark$ & $\checkmark$ & $\checkmark$ & \ding{53} & \ding{53} &$\checkmark$\\
        \hline 
        \cite{10070406}   & \ding{53} & \ding{53} & \ding{53} & \ding{53} & \ding{53} & \ding{53} &\ding{53}\\
        \hline
         \cite{9815180}  & \ding{53} & $\checkmark$  & \ding{53} & \ding{53} & \ding{53} & \ding{53} &\ding{53}\\
        \hline
        \cite{10286997} & $\checkmark$ & $\checkmark$ & \ding{53} & \ding{53} & \ding{53} & \ding{53} &\ding{53}\\
        \hline
        \cite{wen2023}   & \ding{53} & $\checkmark$ & \ding{53} & \ding{53} & \ding{53} & \ding{53} &\ding{53}\\
        \hline
         \cite{10250875} & \ding{53} & \ding{53} & \ding{53} & \ding{53} & \ding{53} & \ding{53} &\ding{53}\\
        \hline
        \cite{9491087} & \ding{53} & \ding{53} & \ding{53} & \ding{53} & \ding{53} & \ding{53} &\ding{53}\\
        \hline
        \cite{kang2024tinymultiagentdrltwins}   & \ding{53} & $\checkmark$ & \ding{53} & \ding{53} & \ding{53} & \ding{53} &\ding{53}\\
        \hline
        \cite{BENJABALLAH2020107099}   & \ding{53} & \ding{53} & $\checkmark$ & $\checkmark$ & \ding{53} & \ding{53} & \ding{53}\\
        \hline
        \cite{luo2023}   & \ding{53} & $\checkmark$  & $\checkmark$ & \ding{53} & $\checkmark$ & $\checkmark$ &\ding{53}\\
        \hline
        \cite{RePEc}  & \ding{53} &\ding{53}  & $\checkmark$ & \ding{53} & $\checkmark$ & \ding{53} &$\checkmark$\\
        \hline
        \cite{10.1007/978-3-031-19211-1_24}   & \ding{53} &\ding{53} & $\checkmark$ & \ding{53} & $\checkmark$ & \ding{53} &$\checkmark$\\
        \hline
        \cite{mobility}  & $\checkmark$ & \ding{53} & \ding{53} & \ding{53} & \ding{53} & \ding{53} &\ding{53}\\
        \hline
        \cite{10234402}  & \ding{53} & \ding{53} & $\checkmark$ & $\checkmark$ & \ding{53} & $\checkmark$ &\ding{53}\\
        \hline
        
        \textcolor{red}{Our work}  & \textcolor{red}{$\checkmark$} & \textcolor{red}{$\checkmark$} & \textcolor{red}{$\checkmark$} & \textcolor{red}{$\checkmark$} & \textcolor{red}{$\checkmark$} & \textcolor{red}{$\checkmark$} &\textcolor{red}{$\checkmark$}\\
        \hline
    \end{tabular}
\end{table*}

In vehicular metaverses, DT technology allows entities such as vehicles and RSUs in the physical space to be accurately mapped and simulated in the virtual space, achieving a high level of integration between virtual and physical spaces. In \cite{10070406}, the authors discussed how users could fully immerse themselves in the virtual space by using VR devices, experiencing visual and auditory effects that transcend reality. In \cite{9815180}, the authors explored the integration of edge intelligence and metaverses, analyzing various challenges faced in applying metaverses under current technological conditions. In \cite{9944868}, the authors proposed the application of mobile edge networks in metaverses and discussed the challenges of realizing edge-enabled metaverses. In \cite{YuezhongMetaverse}, the authors introduced the concept of vehicular metaverses, explained that seamless and immersive services require dynamic migration, and proposed a method based on game theory for twin migration. Besides, the authors in \cite{10286997} combined technologies such as metaverses and DTs to propose the concept of vehicular metaverses, exploring its application in autonomous driving.

\subsection{Twin Migration}

Several studies have explored the facilitation of DT migration. For instance, the authors in \cite{wen2023} proposed a novel metric called Age of Migration Task (AoMT) to measure the freshness of Vehicle Twin (VT) migration tasks. Then, the authors proposed a contract model to motivate RSUs to provide bandwidth resources for VT migration. Since AR tasks require rendering support, which demands substantial computational and storage resources, the authors in \cite{10250875} proposed an efficient resource allocation framework for AR-enabled vehicular edge metaverses. In \cite{9491087}, the authors introduced a DT migration framework for edge networks to reduce DT migration latency. In \cite{kang2024tinymultiagentdrltwins}, the authors proposed a tiny machine learning Stackelberg game framework, which leverages pruning technology for efficient twin migration of unmanned aerial vehicles. However, the above works primarily focus on relatively resource optimization schemes and overlook the critical security challenges associated with DT migration.

\subsection{Network Attacks in Vehicular Networks}

Network attacks are prevalent in both traditional vehicular networks and emerging vehicular metaverses, making security risks a critical concern \cite{BENJABALLAH2020107099}. Some studies have been conducted to defend against network attacks. For instance, the authors in \cite{8293801} presented a real-time DDoS monitoring scheme to estimate the impact of DDoS attacks. However, as attackers continue to develop new techniques and patterns similar to multi-vector attacks, existing monitoring solutions may not be able to detect new and covert attacks. To counteract various covert network attacks, the authors in \cite{10.1007/s10586-023-04035-5} developed a deep learning-based detection mechanism that achieves automatic recognition and real-time response to new attacks by analyzing traffic characteristics and behavior patterns. This mechanism effectively detects DDoS and other complex attacks in Software Defined Network (SDN) environments. Nonetheless, the works mentioned above do not adequately apply to SDN-IoV. To this end, the authors in \cite{9829332} introduced a moving target defense method that dynamically modified network configurations to mitigate DDoS attacks in SDN-IoV. 

Since the existing solutions for mitigating DDoS attacks cannot adapt to dynamic conditions of DT migration in vehicular metaverses \cite{10018506}, the authors in \cite{luo2023} proposed a dual pseudonym scheme and a synchronized pseudonym-changing framework to counter four types of privacy attacks during the VT migration process. Some researchers utilized trust evaluation mechanisms to defend against network attacks in vehicular networks. The authors in \cite{RePEc} proposed a collaborative trust-based method to detect malicious nodes in Vehicular Ad-hoc NETworks (VANETs). To enhance the defense against emerging network attacks, the authors in \cite{10.1007/978-3-031-19211-1_24} developed a privacy-preserving vehicular trust management system using blockchain technology. This system ensures the security of data transactions by analyzing traffic characteristics and behavior patterns, which can achieve automatic recognition and real-time response to new attacks. To improve the security of infotainment services in vehicular networks, the authors in \cite{mobility} proposed a trust-based energy and mobility-aware routing protocol to enhance infotainment services in VANETs. In \cite{kang2024}, the authors proposed a secure and reliable VT migration framework in vehicular metaverses, which designed a two-tier trust evaluation model to comprehensively assess the reputation values of RSUs in both network communication and interaction layers. 

Current defense mechanisms often focus on reactive approaches or address only specific types of attacks. Reactive solutions give attackers more time to probe and exploit system vulnerabilities \cite{luo2023}. In vehicular metaverses, if RSUs are targeted by DDoS attacks, AI agents may be unable to provide real-time services, leading to a significant degradation in service quality \cite{10234402}. 
Furthermore, malicious RSU attacks pose a threat to the integrity of transmitted data, potentially compromising user privacy and decision-making processes. Therefore, it is imperative to develop a proactive security framework to ensure the secure online migration of AI agents within vehicular metaverses. To this end, our work focuses on mitigating DDoS attacks during AI agent migration and evaluating the trustworthiness of RSUs to distinguish malicious RSUs from legitimate ones, thereby providing a comprehensive defense against network threats.

\section{Online AI Agent Migration Framework in Vehicular Metaverses}
\label{sec:sy}

In this section, we introduce the proposed online AI agent migration framework in vehicular metaverses. We first systematically discuss potential network attacks during AI agent migration. We then present the system model and derive the optimization problem for efficient and secure AI agent migration. 
\begin{figure*}[!t]
\centering
\includegraphics[width=0.8\linewidth]{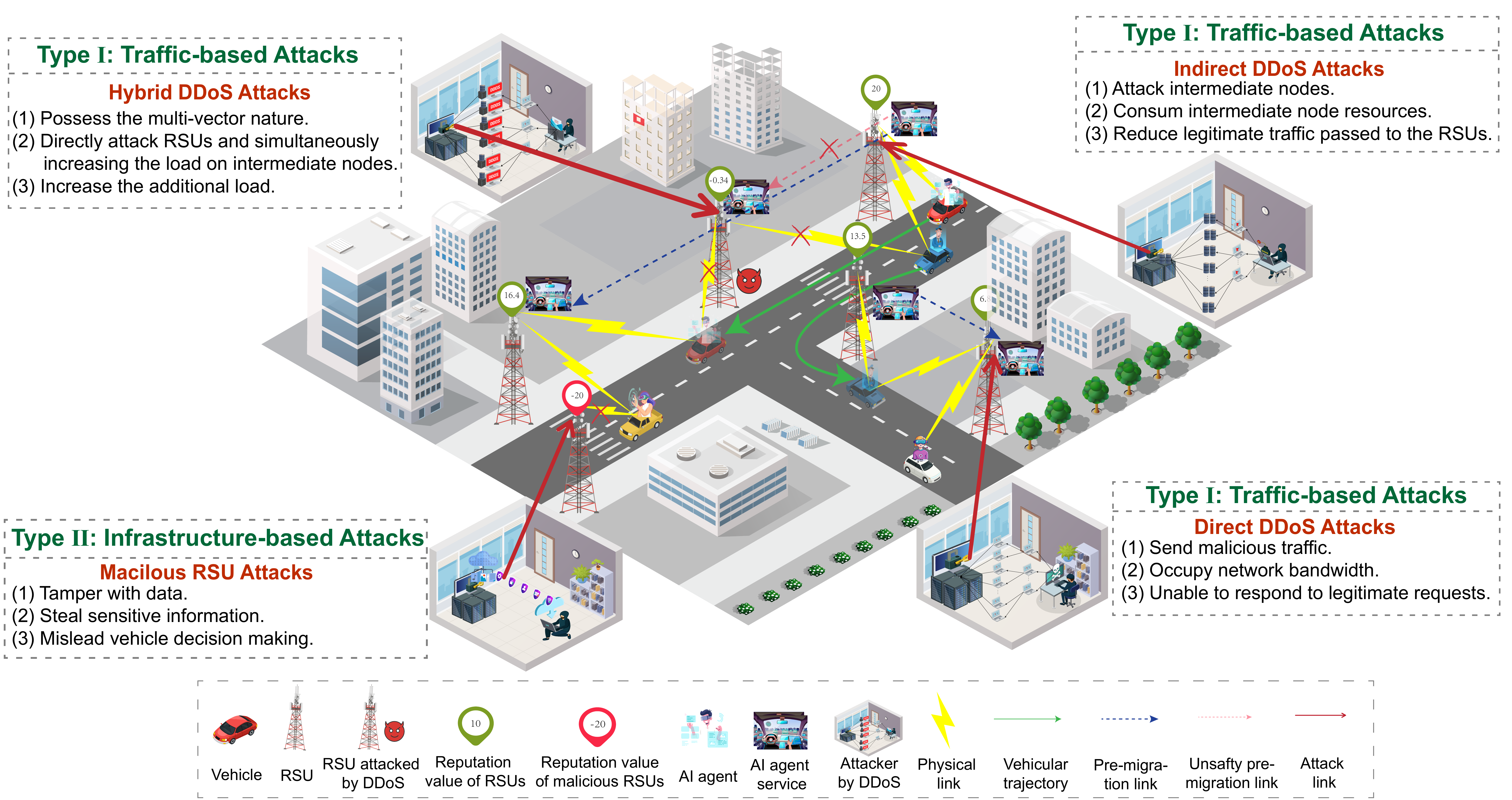}
\caption{Online AI agent migration framework in vehicular metaverses. We systematically present four typical network attacks and their characteristics during AI agent migration. }
\label{fig:env}
\end{figure*}

\subsection{Network Attacks for AI Agent Migration}

During the AI agent migration process, attackers typically employ two types of network attacks against RSUs as shown in Fig. \ref{fig:env}:  \textit{1) Traffic-based Attacks:} They may launch DDoS attacks targeting RSUs based on communication bandwidth or computing workload \cite{10534545}. DDoS attacks are characterized by their stealth and prolonged impact \cite{9829332}. These attacks employ numerous distributed nodes to simultaneously target RSUs, with each node sending minimal data to avoid detection. Although individual data streams may appear innocuous, their cumulative effect can severely strain RSUs, resulting in resource exhaustion, degraded performance, and even network failure \cite{Obaidat2020, 9152970}. Attackers can further obscure their activities by using low-rate data packets that mimic normal network traffic, making it difficult to distinguish malicious traffic from legitimate data. This strategy significantly undermines the effectiveness of traditional DDoS defense mechanisms \cite{10234402}. Such attacks can result in significant service interruptions. \textit{2) Infrastructure-based Attacks:} Attackers may become malicious RSUs that interact with intelligent vehicles, gathering sensitive information such as geographic locations and IP addresses, and potentially altering data to mislead vehicles during the migration decision-making process \cite{8539991}. To amplify the impact of their efforts, the attackers often combine multiple attack strategies. In this paper, we systematically introduce four specific types of network attacks encountered during AI agent migration, specifically,
\begin{enumerate}
\item \textbf{Direct DDoS Attacks:} Direct DDoS attacks involve flooding RSUs with high volumes of malicious traffic, which consumes significant network bandwidth and reduces the capacity available for legitimate communications \cite{jsan12040051}. This attack can overwhelm the network links of RSUs, pushing the load on targeted RSUs to their maximum capacity. As a result, the excessive traffic impedes the ability of RSUs to respond to users, thereby preventing vehicles from accessing crucial information needed for migration decisions.

\item \textbf{Indirect DDoS Attacks:} Indirect DDoS attacks do not target RSUs directly but instead degrade their performance by increasing the traffic load on intermediate nodes, such as routers, servers, and relay nodes. Attackers select these intermediate nodes and flood them with large volumes of fake or malicious traffic, consuming their resources and elevating their loads. When the load on these nodes becomes overwhelming, the capacity for forwarding legitimate traffic to the target RSUs diminishes \cite{10534545}. Consequently, target RSUs receive fewer valid requests, hindering their ability to process AI agents for vehicles in a timely and efficient manner.
 
\item \textbf{Hybrid DDoS Attacks:} Hybrid DDoS attacks, also known as multi-vector DDoS attacks, combine the characteristics of both direct and indirect strategies \cite{10.1007/978-981-99-3481-2_37}. These attacks utilize multiple attack vectors simultaneously, targeting RSUs directly while also increasing the load on intermediate nodes. As a result, target RSUs are burdened with handling direct attack traffic while simultaneously managing the additional network burden caused by the failure of overloaded intermediate nodes. due to the impaired failure of overloaded intermediate nodes. This dual impact amplifies the disruption, making it significantly more difficult for RSUs to maintain effective AI agent services for users.

\item \textbf{Malicious RSU Attacks:} Attackers can deploy numerous attack infrastructures in targeted areas. When RSUs are compromised or owned by attackers, the successfully attacked RSUs become malicious entities. These malicious RSUs can communicate with attack infrastructures to upload collected network information or tamper with private data \cite{8539991}. In vehicular metaverses, RSUs play a pivotal role in the online migration of AI agents. Unlike base stations, RSUs need to support various functions such as real-time data exchange and task migration for intelligent vehicles, handling a large amount of dynamic information. RSUs are more susceptible to hacking, making them a frequent target for attacks \cite{9488287}. Specifically, malicious RSUs can establish connections with vehicles to steal sensitive information such as IP addresses, geographic locations, and sensor data. Additionally, these compromised RSUs can alter vehicle data, delay or disrupt communication, or inject false information, ultimately misleading the decision-making process of vehicles.
\end{enumerate}

\label{section2}
\subsection{System Model}



We consider that the set of vehicles is denoted by $\mathcal{U}  =\left \{ 1,\ldots,u,\ldots, U \right \}$, and the set of RSUs distributed on both sides of the road is denoted by $\mathcal{E} ={\left \{ 1,\ldots,e,\ldots, E \right \}}$, where ${e}$ denotes the RSU within the coverage of current vehicles and ${e^{'}}$ denotes the subsequent RSU. The maximum load capacity of RSUs is denoted by ${L_{e}^{max}}$, and all RSUs have the same coverage area $R_{e}$. Considering a single episode $\mathcal{T} = {\left \{1,\ldots,t,\ldots, T \right \}} $ with $T$ time slots, the coordinate of vehicle ${u}$ is ${P_{u}(t)}=\left [x_{u}(t),y_{u}(t)  \right ] $ and the position coordinate of RSU ${e}$ is ${P_{e}}= \left (x_{e},y_{e}  \right )  $. Firstly, we calculate the latency of AI agent migration to RSUs in vehicular metaverses. Thus, the Euclidean distance between vehicle ${u}$ and RSU ${e}$ at time slot ${t}$ is 
\begin{equation}
  {P_{u,e}(t) }=\sqrt{\left | x_{e} -x_{u}(t) \right |^{2}  +\left| y_{e} -y_{u}(t) \right |^{2}}. 
\end{equation}

Due to the dynamic mobility of vehicles, the distance between vehicles and RSUs is constantly changing, which also affects the wireless communication channel between them. Considering the homogeneity of wireless uplink and downlink channels, the gain ${r_{u,e} (t)} $ of the Rayleigh fading channel among RSUs at time slot ${t}$ is given by \cite{10.114}
\begin{equation}
  {r_{u,e} (t)} =\Lambda \left [ \frac{c}{4\pi fP_{u,e} (t)}   \right ]^{2}, 
\end{equation}
where ${\Lambda} $ is the channel gain coefficient, ${f} $ is the carrier frequency, and ${c}$ is the speed of light. The uplink rate between vehicles and RSUs plays a critical role in determining the time required for vehicles to upload data, including migration requests and the data necessary for service rendering. The transmission latency when vehicle $u$ sends task requests to RSU $e$ is determined by the uplink transmission rate, which can be calculated as follows:
\begin{equation}
  {\nu_{u,e}^{up}(t)  } =B^{up} \log_{2}{\left [  1+\frac{p_{u} r_{u,e}(t) }{\varepsilon _{u}^{2} }\right ]},
\end{equation} 
where ${B^{up}}$ denotes the uplink bandwidth, ${\varepsilon _{u}^{2} }$ denotes the additive Gaussian white noise, and ${p_{u}}$ denotes the transmit power of vehicle ${u}$. Similarly, the downlink rate between vehicle ${u}$ and RSU ${e}$ at time slot ${t}$ can be calculated as
\begin{equation}
 {\nu_{u,e}^{down}(t)  } =B^{down} \log_{2}{\left [  1+\frac{p_{u} r_{u,e}(t) }{\varepsilon _{u}^{2} }\right ]},
\end{equation}
where ${B^{down}}$ denote the downlink bandwidth.





Before AI agent migration, vehicle ${u}$ will upload data including AI agent migration requests with a size of ${S_{u}^{up} (t) } $ to RSU ${e}$, and the uplink latency ${\Gamma _{u,e}^{up} (t) }$ can be calculated as
\begin{equation}
  {\Gamma _{u,e}^{up} (t) } =\frac{S_{u}^{up}(t) }{\nu _{u,e}^{up} (t)}. 
\end{equation}

The size of AI agents is denoted by ${S_{u,e}^{comp} (t)}$. To ensure a seamless experience for vehicular users, vehicle $u$ can pre-migrate a portion of AI agents from the current RSU $e$ to RSU $e^{'}$. As a result, both RSU $e$ and RSU $e^{'}$ can handle decision-making tasks simultaneously \cite{kang2024}. Thus, the result size of AI agent downloaded from RSU ${e}$ is given by
\begin{equation}
  {S_{u,e}^{down} (t)} = S_{u,e}^{comp} (t)-S_{u,e}^{mig} (t).
\end{equation}

At time slot ${t}$, ${S_{u,e}^{mig} (t)}$ can be calculated by \cite{10415630}
\begin{equation}
  {S_{u,e}^{mig} (t)} = \begin{cases}\sigma S_{u,e}^{comp} (t),  & \text{ if }  \gamma_{u}(t)  = 1, \\ 0, & \text{ if }  \gamma_{u}(t)  = 0,
  \end{cases}
\end{equation}
where $\sigma$ is the portion of pre-migrated AI agents and $\gamma_{u}(t) = 1$ represents that vehicle ${u}$ makes AI agent pre-migration decisions, indicating that RSU $e$ pre-migrates some parts of AI agents with a size of ${S_{u,e}^{mig} (t)}$ to the next RSU $e^{'}$ through a wired link.  
Similarly, the result size of AI agent migration from the next RSU ${e{'}}$ is expressed as
\begin{equation}
  {S_{u,e^{'} }^{down}(t)=S_{u,e }^{mig}(t)}.
\end{equation}

Thus, the downlink latency caused by vehicle ${u}$ receiving the feedback of AI agents is given by
\begin{equation}
  {\Gamma _{u,e}^{down} (t) } =\frac{S_{u,e}^{down}(t) }{\nu _{u,e}^{down} (t)} +\frac{S_{u,e^{'} }^{down}(t)}{\nu _{u,e^{'}}^{down} (t)},
\end{equation}
and the latency caused by AI agent pre-migration is \cite{wen2023}
\begin{equation}
  {\Gamma _{u,e}^{mig} (t)=\frac{S_{u,e}^{mig}(t) }{B_{e,e^{'} } } },
\end{equation}
where ${B_{^{e,e^{'}}} }$ represents the bandwidth between RSUs. After vehicle $u$ requests AI agent migration tasks, the tasks must wait for processing at RSU $e$, depending on the current load and computing capacity of RSU $e$. For RSU ${e}$, its processing latency ${\Gamma _{u,e}^{proc}}$ is the time handling AI agent migration, which is expressed as \cite{kang2024}
\begin{equation}
  {\Gamma _{u,e}^{proc}=\frac{L_{e}(t)+S_{u,e}^{comp}-S_{u,e}^{mig}   }{c_{e} }  },
\end{equation}
where ${L_{e}(t)}$ is the initial computation load of RSU ${e}$ at time slot ${t}$ and ${c_{e}}$ stands for the computing power of RSU ${e}$. Similarly, after the pre-migration is completed, the tasks will be processed at RSU $e^{'}$. Therefore, the processing latency ${\Gamma _{u,e^{'}}^{proc}}$ from receiving the request to the completion at RSU $e^{'}$ is expressed as
\begin{equation}
  {\Gamma _{u,e^{'} }^{proc}= \frac{L_{e^{'} }(t)+S_{u,e}^{mig}   }{c_{e^{'} } }   }.
\end{equation}
Thus, the total processing latency of AI agent migration is expressed as
\begin{equation}
  {\Gamma_{u}^{proc} (t)} = \begin{cases}\Gamma_{u,e}^{proc} (t),  & \text{ if }  \Gamma_{u,e}^{proc} (t) < \Gamma_{u,e^{'}}^{proc} (t), \\ \Gamma_{u,e^{'}}^{proc} (t) + \Gamma_{u,e}^{mig} (t), & \text{ otherwise }.
  \end{cases}
\end{equation}


In summary, the total latency ${\Gamma_{u}^{total} (t)}$ of AI agent migration for vehicle $u$ is given by
\begin{equation}
  {\Gamma_{u}^{{total}} (t)} = 
    \Gamma_{u,e}^{{up}}(t) + \Gamma_{u}^{{proc}}(t) + \Gamma_{u,e}^{{down}}(t).
\end{equation}

\section{Problem Formulation}
\label{sec:pf}

During AI agent migration, the utility function of vehicle $u$ involves three factors \cite{9829332}, i.e., the service quantity function $F_Q(t)$, total latency $\Gamma_{u}^{{total}} (t)$, and the security situation function $F_E(t)$. Specifically, the service quantity function is $F_Q(t)=\sum_{e = 1}^E \mathfrak{N}_{u,e}(t)$, where $\mathfrak{N}_{u,e}(t)$ represents the number of vehicles in the set $\mathcal{E}$ connected to RSU ${e}$ at time slot ${t}$, and the security situation function is $F_E(t)=\sum_{e = 1}^E {\mathfrak{N}_{u,e}(t)}{\mathbb{Y}}_{e}(t)$, where ${\mathbb{Y}}_{e}(t)$ is an indicator function, based on DDoS anomaly detection, that takes $1$ if RSU $e$ is attacked, and $0$ otherwise \cite{10015746}. Thus, the utility function of vehicle $u$ for AI agent migration is given by
\begin{equation}
    {U}_u(t) = \lambda F_Q(t)-\beta\Gamma_{u}^{{total}} (t) -\mu F_E(t),
\end{equation}
where $\lambda$, $\beta$, and $\mu$ are coefficients with values greater than $0$. 

To ensure seamless and reliable service experiences for in-vehicle users, it is necessary to determine an optimal pre-migration decision strategy for AI agents, which ensures the security of AI agent migration while minimizing total latency. The optimization problem is formulated as follows:
\begin{align}
\label{formulat:15}
    \max_{\gamma} &\quad  \sum_{t=1}^{T} \sum_{u=1}^{U} {U}_u(t), t \in \{1, \ldots, T\}\\
     \text{s.t.}
     &\quad\text{C1: } L_{e}(t) < L_{e}^{max}, \forall e \in \mathcal{E}\tag{16a} \label{15a}, \\
     & \quad\text{C2: } P_{u,e}(t) < R_{e}, \forall e \in \mathcal{E},\forall u \in \mathcal{U} \tag{16b}, \label{15b}\\
     & \quad\text{C3: } \mathfrak{N}_{u,e}(t) < Q_{u,e}^{max}, \forall e \in \mathcal{E},\forall u \in \mathcal{U} \tag{16c}, \label{15c}\\
     & \quad\text{C4: } Q_{u,e}(t)P_{u,e}(t) < W_{u}(t),\forall e \in \mathcal{E}, \forall u \in \mathcal{U} \tag{16d}, \label{15d}\\
     & \quad\text{C5: } \gamma_{u}(t) \in \{0, 1\}, \forall u \in \mathcal{U}\tag{16e}, \label{15e}\\
     & \quad\text{C6: } \sigma \in [0, 1],\forall u \in \mathcal{U} \tag{16f}, \label{15f}
\end{align}

The constraint C1 ensures that the load \({L_{e}(t)}\) on each RSU does not exceed its maximum handling capacity \({L_{e}^{max}}\). C2 requires that the communication coverage \({R_{e}}\) of the selected RSUs must be greater than their Euclidean distances to the vehicles they serve. C3 specifies that the total number of vehicles connected to RSU $e$  $\mathfrak{N}_{u,e}(t)$ must not exceed its maximum access capacity \({Q_{u,e}^{max}}\). C4 indicates that if a vehicle needs to connect to an RSU but lies outside its coverage range, the connection request must be relayed through neighboring vehicles, where \({Q_{u,e}(t)}\) indicates whether vehicle \({u}\) is connected to RSU \({e}\) at time slot \({t}\) and \({W_{u}(t)}\) represents the maximum communication range of vehicles. C5 specifies whether AI agents need to be pre-migrated. C6 defines the extent of AI agent pre-migration required, where a value of $0$ denotes no pre-migration, values between $0$ and $1$ indicate partial pre-migration, and a value of $1$ represents complete pre-migration. 

The optimization problem (\ref{formulat:15}) involves multiple factors, effectively forming a multi-dimensional knapsack problem that has been proven to be NP-hard \cite{PISINGER20052271}. Additionally, the environmental information during AI agent migration may be incomplete due to the mobility of vehicles \cite{kang2024}. Thus, traditional methods may not be applicable to solving the optimization problem (\ref{formulat:15}). Recently, DRL algorithms have shown significant advantages in obtaining optimal decision-making strategies. Unlike traditional optimization approaches that often require complete and static data, DRL can operate effectively with partial observations of the environment. By leveraging techniques such as POMDPs, DRL algorithms can make informed decisions even when some information is uncertain, a crucial capability for AI agent migration.

\section{MAPPO-based Online AI Agent Pre-Migration Algorithms for DDoS Attacks}
\label{sec:sa}


\subsection{MDP Model}

DRL can learn an optimal policy from experience based on the current state and a given reward without needing prior knowledge of the system. In vehicular metaverses, DRL agents (i.e., vehicles) cannot directly access the complete system state due to communication limitations and the high mobility of vehicles. Instead, they rely on local observations and adaptive strategies to choose actions that maximize cumulative rewards over time. Given the partial observability of network states, we adopt the POMDP to model the problem (\ref{formulat:15}) as follows: 
\begin{enumerate}
    \item [1)]{\textbf{Observation Space:} During the AI agent migration process, AI agents can observe information in the real-time interactive environment between RSUs and vehicles. The set of observation spaces made by vehicles is $\mathcal{O}$, which can be expressed as $\mathcal{O} =\left \{  o_{1}, \dots, o_{u},\dots,o_{U} \right \} $. For vehicle ${u}$, its observation space at time slot ${t}$ is
    \begin{equation}
    {o_{u}^{t} =\left \{ P_{u}(t),L_{e}(t), L_{e^{'}}(t),\Gamma_{u}^{total} (t),\mathrm{\mathbb{J}_{e}(t)}\right \} }, 
    \end{equation}
    where $P_{u}(t)$ represents the real-time location information of intelligent vehicles, ${L_{e}(t)}$ represents the workload on the current RSU ${e}$, ${L_{e^{'}}(t)}$ represents the workload on the next RSU ${e^{'}}$, $\Gamma_{u}^{total} (t)$ is the total latency associated with the migration of AI agents, and $\mathrm{\mathbb{J}_{e}(t)}$ is the frequency of DDoS attacks.
    } 
    \item [2)] {\textbf{Action Space:} For each vehicle, there are only two actions: pre-migrating part of AI agents or directly executing AI agent migration without pre-migration. Specifically, the action space is defined as the pre-migration decisions of vehicles. At time slot \({t}\), the action of vehicle \({u}\) is expressed as \({a_{u}^{t} = \gamma_{u}(t)}\), and the joint actions of all vehicles are represented as \(\mathcal{A} = \{a_{1}, \ldots, a_{u}, \ldots, a_{U}\}\).  
    }
    
    
    \item [3)] {\textbf{Immediate Reward:} At time slot ${t}$, vehicle ${u}$ selects an action ${a}_{u}^{t}$ based on its observation space $o_{u}^{t}$ and receives a reward from the environment. To maximize the utility function $U_u(t)$, the reward function is defined as
    \begin{equation}
    {R}(t) = \sum_{u=1}^{U} {U}_u(t).
    \end{equation}
    }
\end{enumerate}

Our goal is to obtain the optimal AI agent pre-migration strategy while ensuring safety, that is, how to maximize the cumulative reward obtained from the environment. Therefore, we reformulate the optimization problem as follows:
\begin{equation}
\max_{\pi } \mathbb{E}_{\pi } \left [ \sum_{k=0}^{\infty } \Upsilon ^{k} R(t+k)  \right ],
\end{equation}
where $\pi$ is a policy for obtaining the optimal pre-migration strategy, $\mathbb{E}$ is the expectation operator, and ${\Upsilon}$ is a discount factor between $0$ and $1$.

\begin{figure}[t]
\centering
\centerline{\includegraphics[width=0.49\textwidth]{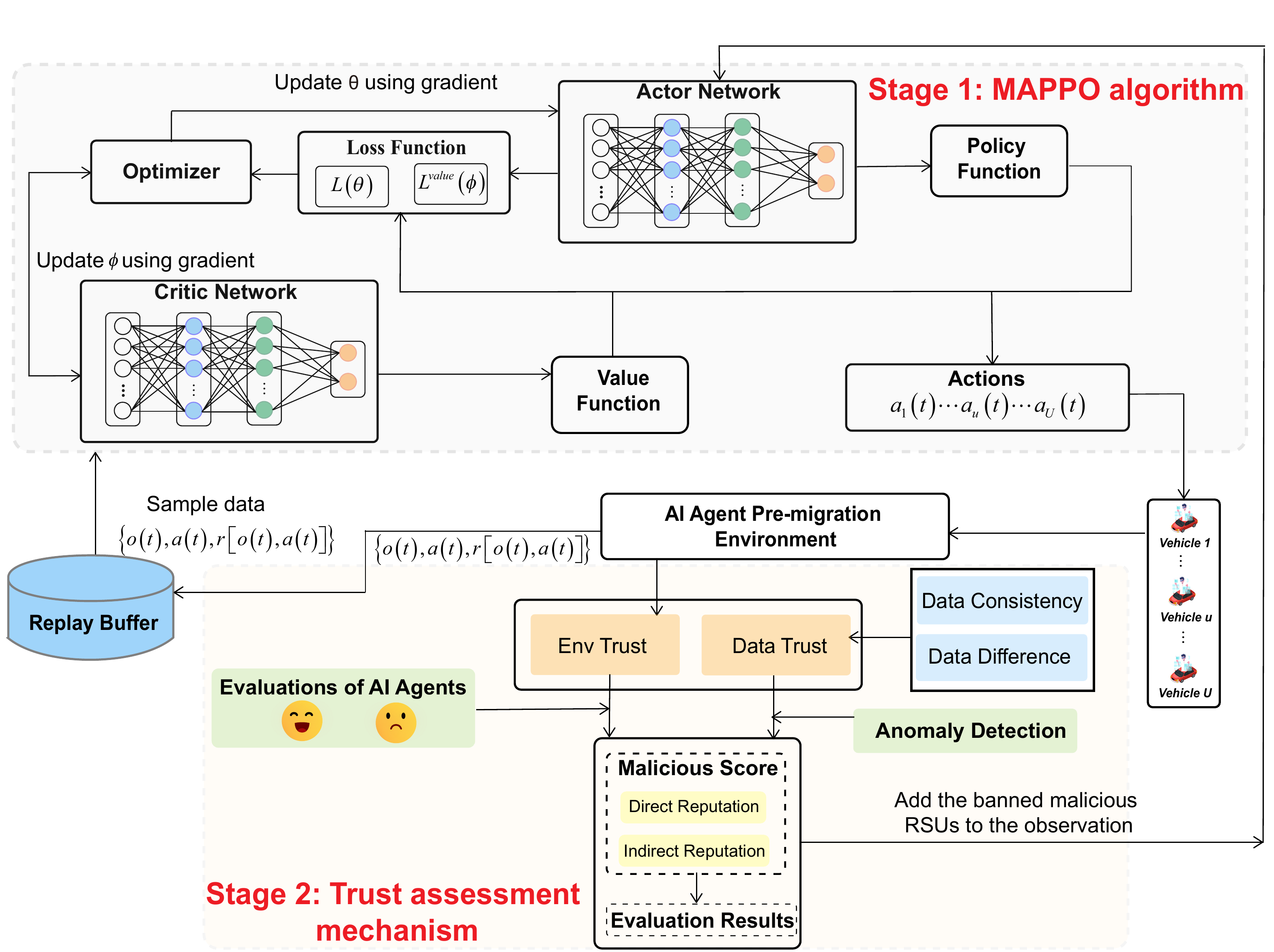}}
\caption{The architecture of the proposed MAPPO-based online AI agent pre-migration algorithm and the trust assessment mechanism.}
\label{fig:alg}
\end{figure}  

\subsection{MAPPO for Optimal Online Migration Strategies}

As the number of vehicles and RSUs in vehicular metaverses continues to grow, the corresponding state and action spaces expand significantly. This increase poses a challenge for algorithms based on Q-Learning, as they lead to a substantial rise in the number of Q-values, resulting in high computational demands and potential overestimation issues. To address these challenges, we adopt the PPO algorithm \cite{9152970}. When extending to scenarios involving collaboration among multiple learning agents, the PPO algorithm can be adapted into the MAPPO algorithm. Unlike the decentralized training and execution framework of independent PPO algorithms, MAPPO utilizes a centralized training and decentralized execution approach \cite{10415630}. Specifically, each agent in MAPPO is equipped with a shared set of parameters and uses aggregated trajectories to update the policy. This method allows all agents to operate under a global policy, enhancing cooperation and achieving superior performance in multi-agent collaborative environments.

MAPPO incorporates the clipping mechanism from PPO to control the magnitude of policy updates, ensuring stable learning and preventing overly aggressive policy shifts \cite{10415630}. For each agent ${u}$, its discrete action policy is defined as ${\pi_{\theta_{u}}}$, which is updated in real time. The objective function for these updates is clipped to effectively regulate policy changes, and the parameters ${\theta_{u}}$ and ${\phi_{u}}$ represent the actor and critic networks, respectively. Thus, the clipped objective function of MAPPO can be formalized as \cite{dewitt2020independentlearningneedstarcraft}
\begin{algorithm}[t]  
    \caption{MAPPO for Optimal Pre-Migration Strategies}  
    \label{alg:MAPPO-OMD}  
    \begin{algorithmic}[1]  
        \STATE Initialize the environment $env$, actor networks $\pi_{\theta_{u}}^{old}$, $\pi_{\theta_{u}}$, the critic network ${L_{\phi_{u}}^{value}}$, the experience replay buffer $\mathcal{B}$, and the batch size $b$;  
        \FOR{$i = 0,1,\ldots, {E}-1$}  
            \FOR{$t = 0,1,\ldots,T - 1$}  
                \FOR{$m = 0,1,\ldots,U-1$} 
                    \STATE Select action ${{a} _{u}^{t} }$ based on $\pi_{\theta_{u}}^{old}$;  
                    \STATE Migrate AI agents to RSUs that are not subject to DDoS attacks;  
                    \STATE Obtain reward ${r}_t$ and next state ${o}_{t+1}$;  
                    \STATE Store $({o}_t, {a}_{t}, {r}_{t}, {o}_{t+1})$ in $\mathcal{B}$; 
                \ENDFOR  
            \ENDFOR  
             \STATE Each vehicle ${u}$ get a trajectory $\chi_{u} = \{{o}_{u}^{t}, {a}_{u}^{t}, {r}_{u}^{t}\}$;
             \STATE Compute ${L_{\phi_{u}}^{value}}$ according to (\ref{critic});
            \STATE Compute $\hat{A}[o_{u}^{t}, a_{u}^{t}]$ using (\ref{GAE});  
            \FOR{$k = 0,1,\ldots, K-1$} 
                \STATE Update $\phi_{u}$ according to (\ref{uc}); 
                \STATE Update $\theta_{u}$ according to (\ref{ua}) ;  
            \ENDFOR
            \STATE Update $\theta_{u}^{old} \gets \theta_{u}$; 
            \STATE Update $\phi_{u}^{old} \gets \phi_{u}$; 
            \STATE Update $\pi_{\theta_{u}}^{old} \gets \pi_{\theta_{u}}$; 
        \ENDFOR  
    \end{algorithmic}  
\end{algorithm}

\begin{equation}
\label{clip}
\begin{split}
{L}^{clip}(\theta_{u}) = \mathbb{E}_t \Big[ \min \Big( 
    & \, r_{t}(\theta_{u}) \hat{A}[o_{u}^{t}, a_{u}^{t}], \\
    & \, {f}_{clip}[r_{t}(\theta_{u}] \hat{A}[o_{u}^{t}, a_{u}^{t}] 
\Big) \Big]
\end{split}
\end{equation}
where \(r_{t}(\theta_{u})\) represents the ratio of the new policy to the old policy, which is the probability ratio of taking action \(a_{u}^{t}\) under the current policy versus the old policy in state \(o_{u}^{t}\), and can be calculated as
\begin{equation} 
\label{ratio}
      r_{t}(\theta_{u}) = \frac{\pi_{\theta_{u}}(a_{u}^{t} | o_{u}^{t})}{\pi_{\theta_{u}}^{old}(a_{u}^{t} | o_{u}^{t})}.
\end{equation}

The clip function operation is used to restrict the ratio \(r_{t}(\theta_{u})\) to the range \([1 - \epsilon, 1 + \epsilon]\), which indicates that if \(r_{t}(\theta_{u})\) exceeds this range, it will be clipped to the boundary value of \(1 - \epsilon\) or \(1 + \epsilon\). The specific formula is given by
\begin{equation}
     f_{clip}[r_{t}(\theta_{u})]= 
     \begin{cases} 
        1 - \epsilon, & \text{if } r_{t}(\theta_{u}) < 1 - \epsilon, \\
        1 + \epsilon, & \text{if } r_{t}(\theta_{u}) > 1 + \epsilon, \\
        r_{t}(\theta_{u}), & \text{otherwise}.
\end{cases}
\end{equation}

\(\hat{A}[o_{u}^{t}, a_{u}^{t}]\)  represents the joint advantage function used to evaluate the quality of actions taken by agent $u$ given its observations. When evaluating policies, MAPPO utilizes \(\hat{A}[o_{u}^{t}, a_{u}^{t}]\) to gauge the advantage of the current policy, ensuring that each update effectively reduces variance in policy estimation, which can be calculated as \cite{10415630}
\begin{equation} 
\label{GAE}
\hat{A}[o_{u}^{t}, a_{u}^{t}] = Q[o_{u}^{t}, a_{u}^{t}] - \frac{1}{|A|} \sum_{a_{u}^{t} \in A} Q[o_{u}^{t}, a_{u}^{t}],
\end{equation}
where \( Q[o_{u}^{t}, a_{u}^{t}] \) represents the Q-value for executing action \( a_{u}^{t} \) under the specific observation \( o_{u}^{t} \), while \( |A| \) denotes the total number of possible actions. \( Q[o_{u}^{t}, a_{u}^{t}] \) represents the expected cumulative reward that can be obtained by starting from observation \( o_{u}^{t}\) and action \( a_{u}^{t} \) while following the current policy, which can be given by \cite{dewitt2020independentlearningneedstarcraft}
\begin{equation}
    Q[o_{u}^{t}, a_{u}^{t}] = \mathbb{E} \left[ \sum_{k=0}^{\infty} \gamma^k r(t+k) \, | \, o_{u}^{t}, a_{u}^{t} \right].
\end{equation}

The $Q[o_{u}^{t}, a_{u}^{t}]$ function is used to evaluate the actions taken by all agents and guides them toward selecting more favorable actions. The critic network is parameterized by \(\phi_u\) and updated using gradient descent with a loss function
\begin{equation}
\label{critic}
      {L}^{value}(\phi_u) = \mathbb{E}_t \left[ (V_{\phi_u}(o_{u}^{t}) - r(t))^2 \right],
\end{equation}
where \(r(t)\) is the target value, ${V_{\phi_u}(o_{u}^{t})}$ represents the expected return that agent $u$ can achieve in the future when starting from ${o_{u}^{t}}$, and ${\phi_{u}}$ represents the set of parameters for the critic network. The critic network updates its parameters by minimizing the loss, given by
\begin{equation}
\label{uc}
    \phi_{u} \leftarrow \phi_{u} - \alpha \nabla_{\phi_u} {L}^{value}(\phi_u),
\end{equation}
where \(\alpha\) is the learning rate. The update of the actor network depends on comparing the current policy with the previous one to ensure stability during updates. Once the critic network provides an accurate value estimate, the actor network leverages this information to refine its policy. The parameters of the actor network are updated using the following policy gradient equation:
\begin{equation}
\label{ua}
    \theta_{u} \leftarrow \theta_{u} + \alpha \nabla_{\theta_{u}} {L}^{total}(\theta_{u}),
\end{equation}
where \(\theta\) represents the parameters of the actor network and \({L}^{total}(\theta_{u})\) denotes the overall loss function, which can be expressed as \cite{dewitt2020independentlearningneedstarcraft}
\begin{equation}
\begin{split}
{L}^{total}(\theta_u) = & c_1 {L}^{value}(\phi_u) + {L}^{clip}(\theta_u)\\
&   - c_2 \mathbb{E}_t \left[ {H}(\pi_{\theta_u}(a_u^t | o_u^t)) \right],
\end{split}
\end{equation}
where $c_1$ and $c_2$ are hyperparameters that control the relative importance of different loss components and $H(\pi_{\theta_{u}}( a_{u}^{t} | o_{u}^{t}))$ is the entropy loss, which encourages exploration by penalizing deterministic policies.

The pseudocode of MAPPO-based online AI agent pre-migration algorithms is presented in Algorithm \ref{alg:MAPPO-OMD}. Initially, the environment, parameters, and replay buffer for AI agent migration are set up. The loop of the algorithm begins at line 2, where interaction with the environment generates samples (lines 2-10), iterating from an initial state until completing $T$ time slots, referred to as an episode. During each episode, the algorithm observes the current network environment, executes the policy $\pi_{\theta_{u}}^{old}$, and selects an action aimed at facilitating secure AI agent migration while mitigating threats from DDoS attacks. The rewards are recorded, and state transition samples are stored in the replay buffer (lines 7-8). Advantage and value functions are estimated using joint advantage functions, and the clipped objective function and value function are updated (lines 12-13). Subsequently, the actor network parameters and critic network parameters are updated (lines 15-16). Finally, the policy is updated (lines 18-20). The computational complexity of the proposed algorithm is expressed as $O(ETU(U+K))$ \cite{10415630}, where ${E}$ represents the number of training episodes, $T$ denotes the maximum number of steps per training episode, $K$ signifies the number of updates required for the policy and value networks following trajectory collection, and ${U}$ signifies the number of agents.

The MAPPO-based online AI agent pre-migration algorithm empowers agents to efficiently select optimal actions based on real-time environmental observations. By employing this continuous learning strategy, the mechanism facilitates AI agent migration to secure RSUs, thus mitigating the impact of DDoS attacks and significantly reducing their adverse effects. However, in edge computing environments, DDoS attacks may represent only one aspect of broader network threats, as multiple types of attacks can be launched concurrently, which means RSUs may be subjected to various other threats, rendering them potentially malicious. To differentiate between normal and malicious RSUs, we propose a reputation-based mechanism aimed at maximizing defensive capabilities and strengthening resilience against adversarial attacks during AI agent migration.

\section{Trust Assessment Mechanism for Malicious RSU Attacks}
\label{sec:ta}

In this section, we propose a trust assessment mechanism to resist malicious RSU attacks, which distinguishes malicious RSUs by assessing their trustworthiness, as shown in Fig. \ref{fig:alg}. 

Malicious RSUs, driven by their intent to steal data or disrupt communications, initiate connection requests with nearby vehicles more frequently than normal RSUs to gain access to more vehicle data or manipulate information \cite{9488287}. In contrast, vehicles assess RSUs based on their trustworthiness. Over time, this assessment process causes a divergence in the trust levels between normal and malicious RSUs, ultimately exposing the malicious ones. This assessment requires consideration of multiple constraints and continuous monitoring of RSUs based on their malicious scores. Malicious RSUs are eventually restricted from associating with vehicles.

\subsection{Trust Assessment}
To achieve RSU trust assessment, we calculate malicious scores of RSUs by considering both direct and indirect factors \cite{siddiqua}. Firstly, direct factors can dynamically reflect the direct interactions between RSUs and vehicles. Indirect factors involve real-time monitoring of packet reception, packet loss rate, and retransmission count of RSUs, which can indirectly assess whether the RSU exhibits malicious behavior. Combining these two factors helps to comprehensively evaluate the RSU's trustworthiness and accurately identify malicious RSUs. Specifically,
\begin{enumerate}
\item \textbf{Direct Factors:} RSUs are directly monitored for abnormal changes in packet transmission, including data modification, duplication, theft, and tampering. If the number of abnormal packets sent by RSU $e$ exceeds a pre-defined threshold, the target RSU ${e}$ is classified as having been maliciously attacked, expressed as \cite{RePEc}
\begin{equation}
\mathbb{D}_{e}(t) = 
\begin{cases} 
    1, & \text{if} \ {N_{e}^{abn}(t)} > {N_{e}^{thr}(t)}, \\
    0, & \text{otherwise},
\end{cases}
\end{equation}
where $N_{e}^{abn}(t)$ represents the number of abnormal packets sent by RSU $e$ at time slot $t$. $\mathbb{D}_{e}(t) = 1$ indicates that the number of detected abnormal packets exceeds the threshold, suggesting that RSU ${e}$ is under malicious attacks. Conversely, the target RSU $e$ is normal. Thus, the direct malicious score of RSU $e$ can be expressed as
\begin{equation}
X_{e}(t) = \sum_{\tilde{t}=1}^{t}{\mathbb{D}_{e}(\tilde{t})}.
\end{equation}

\item \textbf{Indirect Factors:} We consider the completion rate of RSUs in processing decision-making tasks of AI agent migration as an indirect factor for calculating malicious scores. Specifically, the rate at which RSUs complete AI agent migration is determined by AI agents based on the evaluation of total migration latency. If the total migration latency falls below the acceptable latency threshold, AI agents provide positive feedback to RSU $e$, incrementing the completion count ${\omega_{u,e}^{num}}$ by $1$. Otherwise, negative feedback is given. Consequently, the completion count for RSU $e$ at time slot $t$ can be expressed as
\begin{equation}
\omega_{u,e}^{num} (t) =   
\begin{cases}   
    \sum\limits_{\tilde{t}=1}^{t}{\omega_{u,e}^{num} (\tilde{t}) + 1}, & \text{if } {\Gamma_{u}^{total}}(t) \leq {\Gamma_{u}^{thr}}, \\   
    \sum\limits_{\tilde{t}=1}^{t}\omega_{u,e}^{num} (\tilde{t}), & \text{otherwise},  
\end{cases} 
\end{equation}
where ${\Gamma_{u}^{total}}(t)$ represents the actual total migration latency experienced by RSU $e$ at time slot $t$, and ${\Gamma_{u}^{thr}}$ is the latency threshold acceptable to vehicle $u$. Thus, the completion rate ${\xi_{u,e} (t)}$ is expressed as
\begin{equation}
 {\xi_{u,e} (t)= \frac{\omega_{u,e}^{num} (t) }{\omega_{u,e}^{total} (t) } },
\end{equation}
where ${\omega_{u,e}^{total} (t)}$ is the total number of AI agent migrations performed on RSU ${e}$. If ${\xi_{u,e} (t)}$ is significantly lower than other RSUs in the same region and cannot be attributed to external factors such as network congestion or hardware failures, RSU $e$ may be a potential malicious RSU and attempts to disrupt AI agent migration. If ${\xi_{u,e} (t)}$ demonstrates abnormally large fluctuations, such as maintaining a consistently normal performance but suddenly dropping sharply during a specific time slot, it may indicate that this RSU is employing covert strategies to evade detection and conceal malicious behavior. Therefore, ${\xi_{u,e} (t)}$ serves as a critical metric for assessing whether an RSU might be malicious.
\end{enumerate}

\begin{algorithm}[t]
    \caption{Trust Assessment Mechanism}
    \label{alg:RSU_Malicious_Rating}
    \begin{algorithmic}[1]
        \STATE Initialize malicious scores of RSUs to $0$;
        \FOR{$i = 0,1,\ldots, {E}-1$}  
            \FOR{$t = 0,1,\ldots, {T}-1$}
                \FOR{$m = 0,1,\ldots, {U}-1$} 
                    \STATE Execute the action $a_u^t$ based on Algorithm \ref{alg:MAPPO-OMD};
                    \STATE Obtain the completion count $\omega_{u,e}^{num} (t)$ for RSU $e$ from Algorithm \ref{alg:MAPPO-OMD}; 
                    \STATE Update the completion rate ${\xi_{u,e} (t)}$. 
                \ENDFOR
                \STATE Update malicious scores $\mathcal{W}_{e}(t)$ based on (\ref{mal});
                \IF{${\mathcal{W}_{e}(t)}$ > ${Y}$}
                    \STATE RSU $e$ is considered malicious and will be banned.
                \ENDIF
                \STATE Update the malicious threshold ${Y}$ based on (\ref{upd}).
            \ENDFOR
        \ENDFOR
    \end{algorithmic}
\end{algorithm}

The above direct and indirect factors can be obtained from the historical data and the environment. We define the malicious score of RSUs at time slot $t$ as $\mathcal{W}_{e}(t)$, which is used to assess whether RSUs are safe and reliable for AI agent migration. Thus, the malicious score of RSUs at time slot ${t}$ is given by 
\begin{equation}
\label{mal}
{\mathcal{W}_{e}(t) } =\vartheta\sum_{\tilde{t}=1}^{t}{\mathbb{D}_{e}(\tilde{t})  }- {\varrho \sum_{\tilde{t}=1}^{t}\mathfrak{N}_{u,e}(\tilde{t})  \xi_{u,e} (\tilde{t}) } -{\varpi  \sum_{\tilde{t}=1}^{t}\tilde{t}},
\end{equation}
where $\varrho$, $\vartheta$, and $\varpi$ are coefficients with values greater than $0$. The third term $\sum_{\tilde{t}=1}^{t}\tilde{t}$ indicates that the score of malicious scores will decrease over time. When ${\mathcal{W}_{e}(t) }$ is less than the threshold ${Y} $, RSUs will be considered malicious and prohibited from performing AI agent migration.

\begin{table}[t]   
    \centering  
    \caption{The key parameters in the experiment}  
    \resizebox{\columnwidth}{!}{  
        \begin{tabular}{p{5.2cm} c} 
            \hline  
            \textbf{Description} & \textbf{Values} \\
            \hline  
            Number of vehicles ($U$) & $20$ \\
            Number of RSUs ($E$) & $15$ \\
            Time slot ($T$) & $20$ s\\
            Computation capacity range of all RSUs ($c_{e}$) & $[100, 300]$ MHz \cite{10415630}\\
            Size range of AI agents uploaded by all vehicles ($S_{u}^{up}$) & $[25, 200]$ MB \cite{10415630} \\
             Migration bandwidth in RSUs ($B_e$) & $[0.25, 2]$ Gbps \cite{10415630}\\
             Number of episodes & $10^3$ \\
             Learning rate ($\alpha$) & $10^{-4}$ \\
             Discount factor (${\Upsilon}$) & $0.95$ \\
             Batch size ($b$) & $200$ \\
            Clip parameter & $0.05$ \\
              Maximum capacity of the replay buffer ($B$) & $2 \times 10^6$ \\
            \hline  
        \end{tabular}  
    }  
    \label{sec:tab}
\end{table}

\begin{figure*}[t]
    \begin{center}
	\begin{minipage}[t]{0.45\linewidth}
		\centering
            \includegraphics[width=1\textwidth]{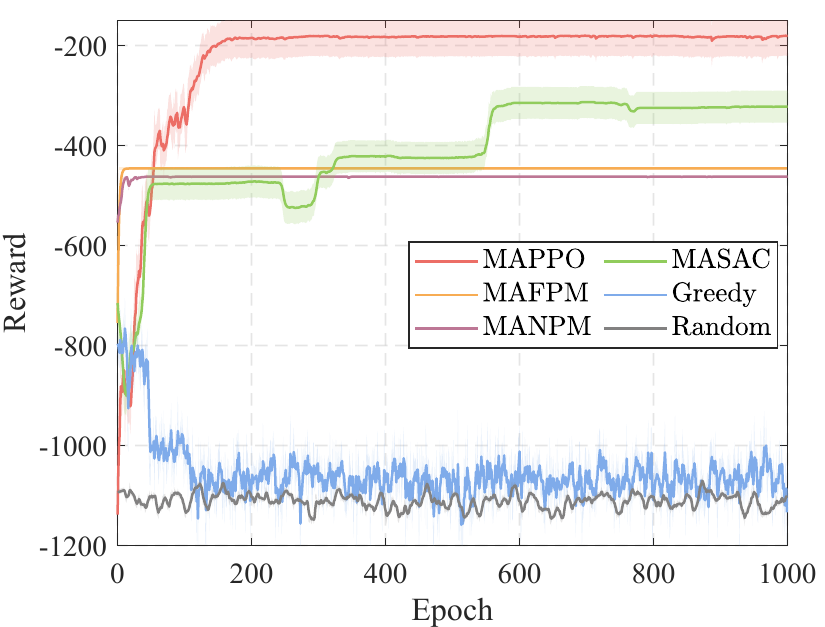}
            \caption{Average test rewards of MAPPO and other baseline algorithms.}\label{fig:Reward}
	\end{minipage}
	\hspace{0.5in}
	\begin{minipage}[t]{0.45\linewidth}
		\centering
            \includegraphics[width=1\linewidth]{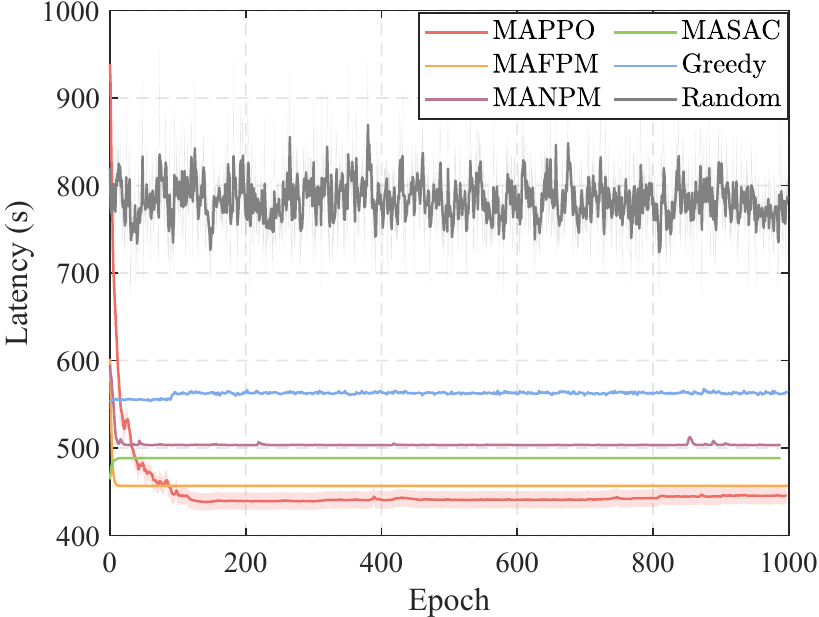}
            \caption{Average total migration latency under MAPPO and other baseline algorithms.}\label{fig:latency}
	\end{minipage}
	\end{center}
\end{figure*}

\begin{figure*}[t]
    \begin{center}
	\begin{minipage}[t]{0.45\linewidth}
		\centering
		\includegraphics[width=1\linewidth]{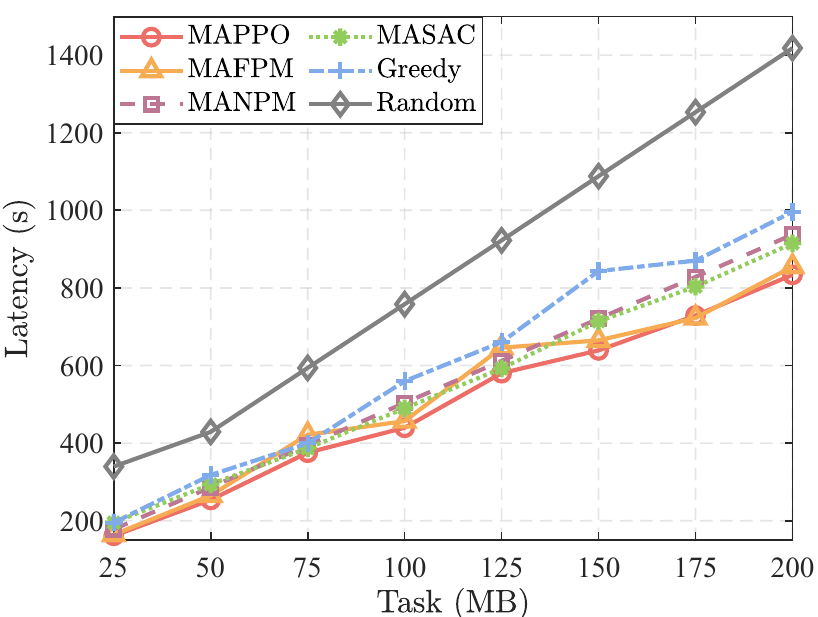}
		\caption{Average total migration latency under different sizes of AI agent migration tasks.}\label{fig:task}  
	\end{minipage}
	\hspace{0.5in}
	\begin{minipage}[t]{0.45\linewidth}
		\centering
            \includegraphics[width=1\linewidth]{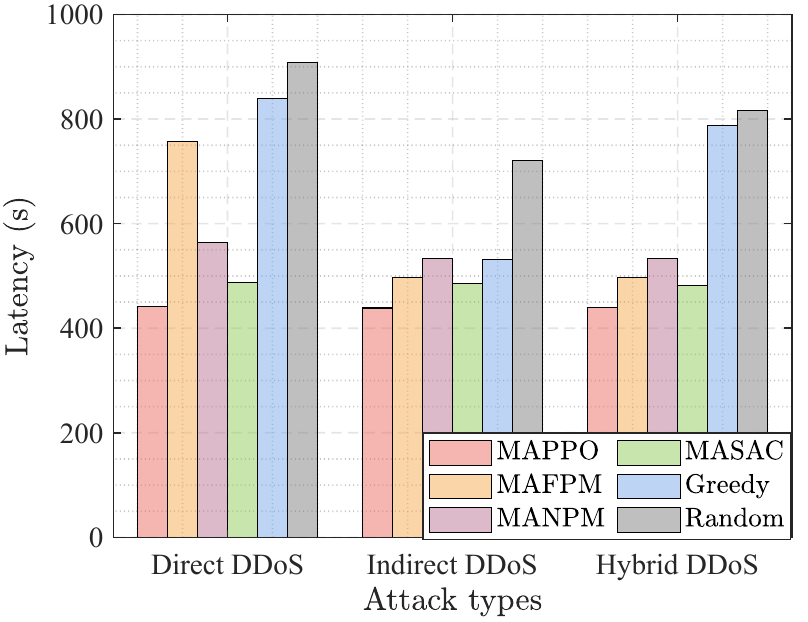}
            \caption{Average total migration latency under different types of DDoS attacks.}\label{fig:type}
	\end{minipage}
	\end{center}
\end{figure*}
\subsection{Adaptive Threshold for Malicious RSUs}

To better adapt to ever-changing attack patterns based on the past behavior of malicious RSUs and current conditions, we consider an adaptive threshold variation, which employs a threshold adjustment mechanism to identify malicious RSUs \cite{KERRACHE20181}. RSUs start with an initial malicious score of $0$, and the threshold is set to \(\varsigma\), where \(\varsigma\) varies within the range of $[0, 1]$. This approach prevents the system from overreacting while gradually adapting to different attack patterns and network conditions. Moreover, the threshold is adaptively adjusted based on changes in the malicious scores of RSUs. The adaptive threshold is given by
\begin{equation}
\label{upd}
    {Y}_{{new}} =
    \begin{cases} 
    {Y}_{{old}} + \varsigma, & \text{if } \mathrm{BR} \leq \tau, \\
    {Y}_{{old}} - \varsigma, & \text{if } \mathrm{FBR} > \zeta, \\
    {Y}_{{old}}, & \text{otherwise},
    \end{cases}
\end{equation}
where $\tau$ and $\zeta$ represent the value thresholds of the False Banning Rate (FBR) and the Banning Rate (BR), respectively, ranging from $0$ to $1$. The FBR represents the probability that a normal RSU is mistakenly identified as a malicious RSU, and the BR refers to the proportion of successfully identified and banned malicious RSUs to the total number of malicious RSUs \cite{9829332}. We have,
\begin{equation}
\mathrm{FBR} = \frac{\mathrm{FP}}{\mathrm{FP} + \mathrm{TN}},
\end{equation}
\begin{equation}
\mathrm{BR} = \frac{\mathrm{TP}}{\mathrm{TP} + \mathrm{FN}},
\end{equation}
where $\mathrm{TP}$ represents the number of malicious RSUs that are correctly identified as malicious RSUs. $\mathrm{TN}$ represents the number of normal RSUs that are correctly identified as normal RSUs. $\mathrm{FP}$ represents the number of normal RSUs that are misclassified as malicious RSUs. $\mathrm{FN}$ represents the number of malicious RSUs that are misclassified as normal RSUs \cite{mobility}.

The pseudo-code of the proposed trust assessment mechanism is illustrated in Algorithm \ref{alg:RSU_Malicious_Rating}. Initially, the malicious score of RSUs is initialized (line 1). At each time slot, actions are executed according to Algorithm \ref{alg:MAPPO-OMD} (line 5). Then, the completion count $\omega_{u,e}^{num} (t)$ is obtained from Algorithm \ref{alg:MAPPO-OMD} and the completion rate ${\xi_{u,e} (t)}$ is updated  (lines 6-7). Then, malicious scores of all RSUs are updated (line 9). If the malicious score of an RSU exceeds the malicious threshold, the RSU is prohibited from connecting to any vehicle (lines 10-12). Finally, the malicious threshold is updated (line 13). The computational complexity of the proposed trust assessment mechanism is ${O}(ETU^2)$ \cite{YuezhongMetaverse}.
\section{Numerical Results} \label{sec:e}
\subsection{Parameter Settings}

We simulate a scenario with $15$ RSUs and $20$ vehicles on a $\SI{10}{\square\kilo\meter}$ urban arterial road, with the time slot of the environment set to $20$ seconds. Each RSU in the environment is equipped with $2.2$ GHz computational capability and is interconnected via a wired network \cite{10415630}. All RSUs are evenly distributed on one side of the road, and every vehicle is within the coverage of at least one RSU, traveling at the same speed. In this environment, we set the size of uploaded data by vehicles to $20$ MB, with the task size of AI agent migration ranging from $25$ MB to $200$ MB. Additionally, the initial bandwidth between vehicles and RSUs is established within a range of $100$ MHz to $200$ MHz. The key parameters of the experiment are detailed in Table \ref{sec:tab}.
\subsection{Convergence Analysis}
\begin{figure*}[t]
\centering
\subfigure[Direct DDoS attacks.]{
\centering
\includegraphics[width=0.3\textwidth]{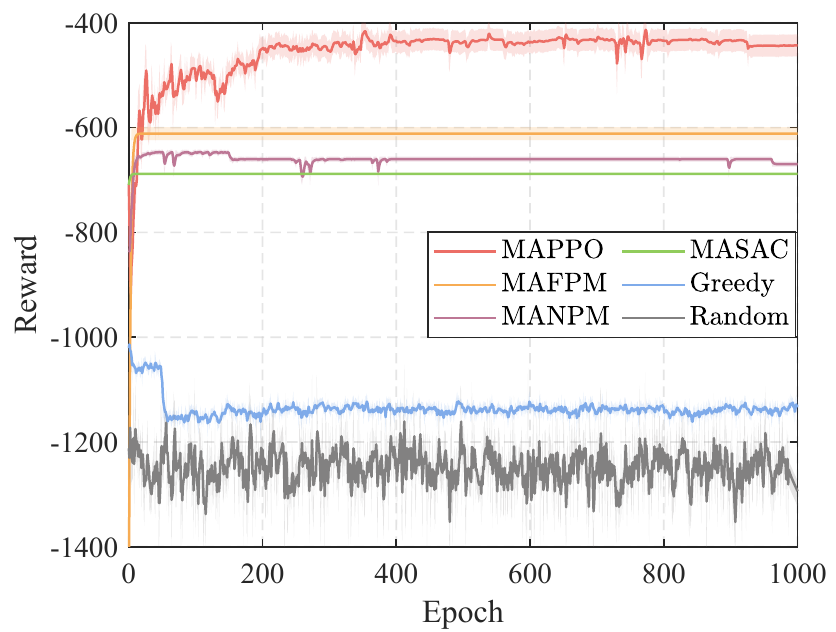}
\label{dir}
}
\subfigure[Indirect DDoS attacks.]{
\centering
\includegraphics[width=0.3\textwidth]{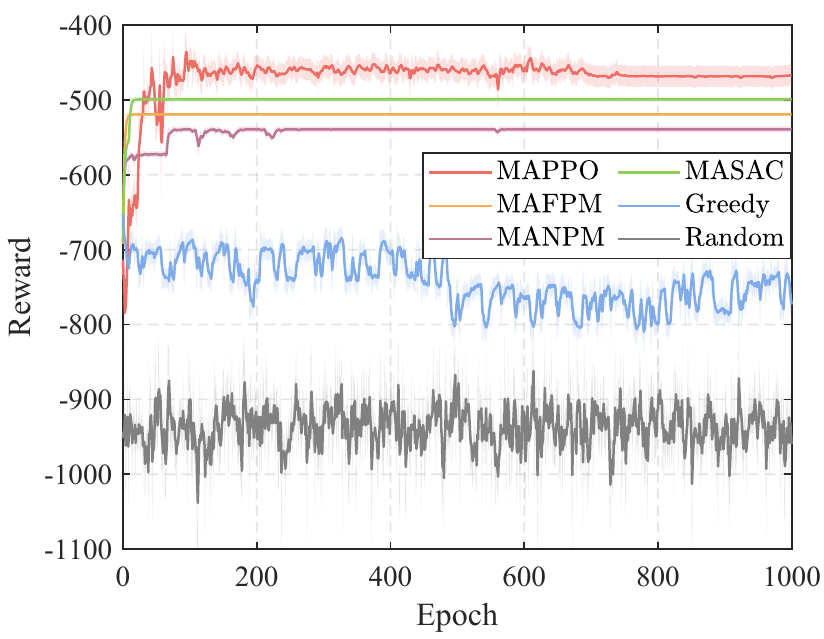}
\label{indir}
}
\subfigure[Hybrid DDoS attacks.]{
\centering
\includegraphics[width=0.3\textwidth]{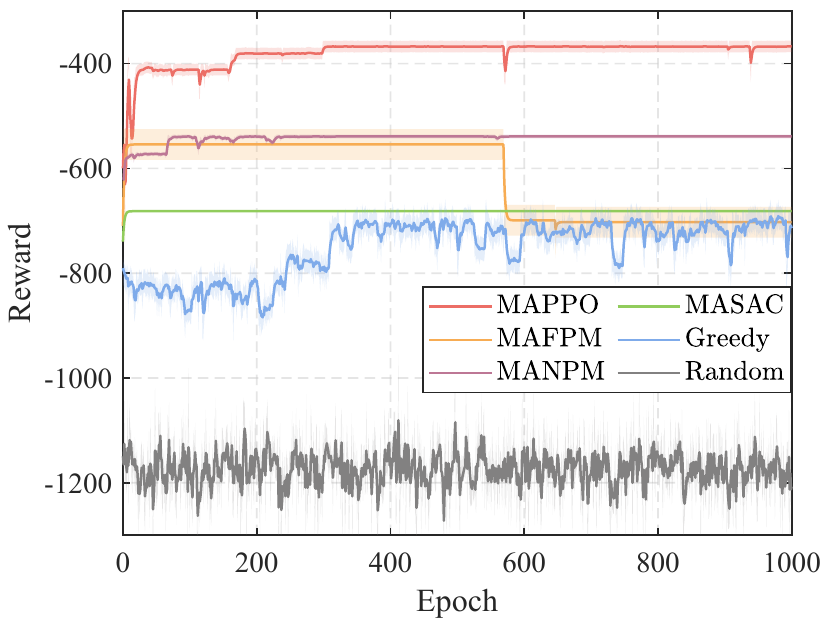}
\label{hyb}
}
\caption{Average test rewards of MAPPO and other baseline algorithms under different DDoS attack types.}
\label{Attacks}
\end{figure*}

\begin{figure*}[t]
\centering
\subfigure[False banning rate.]{
\centering
\includegraphics[width=0.45\textwidth]{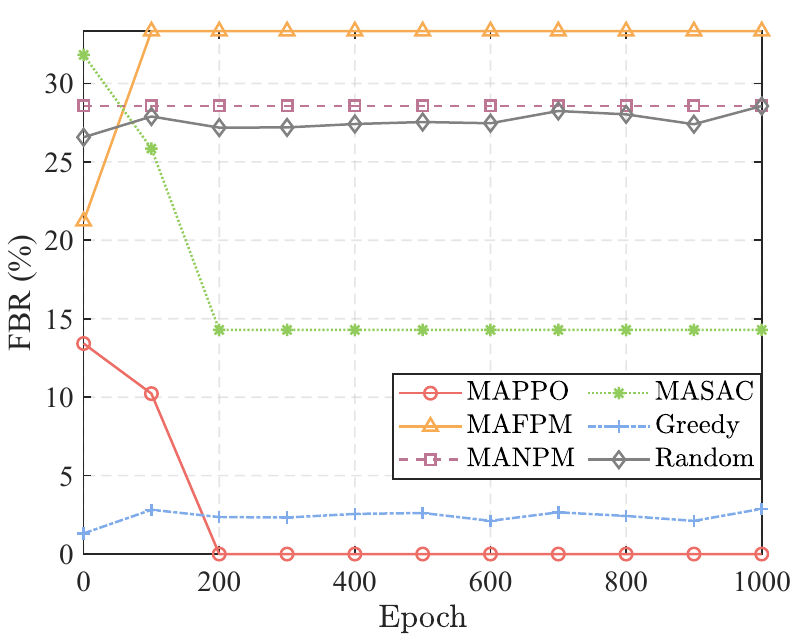}
\label{fbr}
}
\subfigure[Banning rate.]{
\centering
\includegraphics[width=0.45\textwidth]{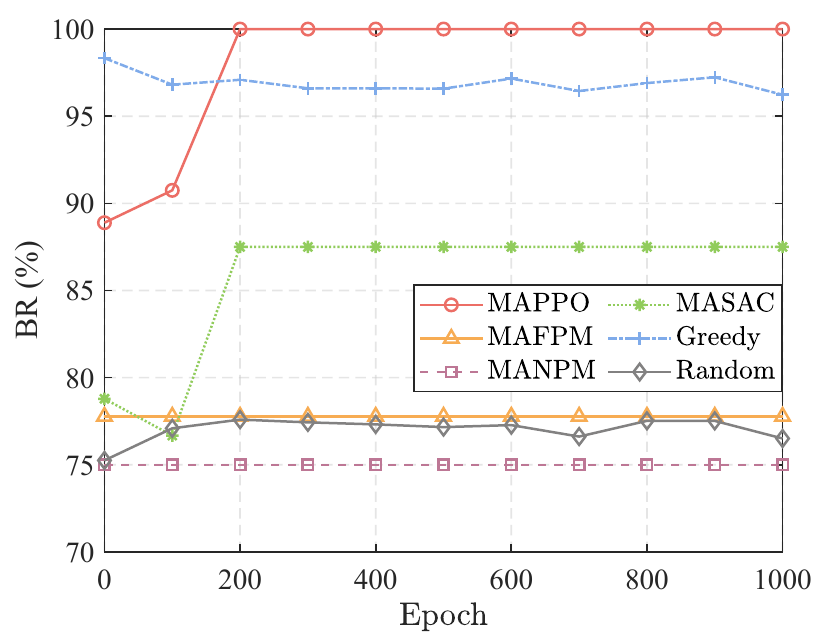}
\label{br}
}
\caption{Performance evaluation of the proposed trust assessment mechanism.}
\label{Trustworthiness performance}
\end{figure*}
To validate the convergence of our proposed MAPPO algorithm, we design several baseline algorithms for comparison, including Multi-Agent Full Pre-migration (MAFPM), Multi-Agent No Pre-migration (MANPM), Multi-Agent Soft-Actor Critic (MASAC), Greedy, and Random algorithms \cite{10415630}. MAFPM represents full pre-migration of tasks, while MANPM refers to scenarios where AI agents are not pre-migrated. As shown in Fig. \ref{fig:Reward}, the proposed algorithm significantly outperforms these baselines. Specifically, our solution shows improvements that of $60.5\%$, $62.2\%$, $76.97\%$, $81.8\%$, and $83.3\%$ compared with MAFPM, MANPM, MASAC, Greedy, and Random, respectively. The Greedy solution performs poorly compared with DRL-based methods, as it only maximizes short-term rewards and neglects long-term benefits. Besides, the poor performance of MAFPM and MANPM stems from increased unnecessary latency, full pre-migration introduces extra migration latency, while AI agent migration without performing pre-migration strategies often faces limited RSU computing resources, leading to increased processing latency. By contrast, the proposed MAPPO solution can process environmental observations in real time, even under high-speed vehicle movement. The reason is that MAPPO allows agents to share local information, effectively dealing with complex environments that may involve network attacks. Consequently, our proposed algorithm demonstrates its effectiveness in solving the secure online migration of AI agents, with significantly higher stability after convergence.

As shown in Fig. \ref{fig:latency}, the proposed algorithm achieves significantly lower average total latency of AI agent migrations for vehicles compared with the baseline methods. The average latency of each vehicle decreases as the number of training steps increases. Specifically, our approach reduces latency by up to $43.4\%$ compared with the baseline solutions. The performance of several baseline algorithms shows relatively high latency, which can be attributed to the decision-making process of vehicles impacting the RSU load in different environments. For example, AI agent migration without performing pre-migration strategies tends to increase the load on RSUs, leading to higher processing latency. Additionally, the downlink transmission rate is closely related to the distance between RSUs and vehicles. When the current RSU has sufficient resources to efficiently process the service requests uploaded by vehicles, adopting a full pre-migration strategy may result in unnecessary downlink transmission delays and an increase in overall pre-migration latency. This is because, in such cases, preemptively migrating large volumes of data not only consumes limited bandwidth resources but may also lead to bandwidth bottlenecks, thereby negatively impacting the overall system performance and ultimately causing a significant increase in total latency. In contrast, our proposed algorithm, MAPPO, adopts pre-migration strategies in response to DDoS attacks, providing more migration options and dynamically adjusting the proportion of AI agent pre-migration, effectively mitigating the impact of attackers. 


\subsection{Performance on AI Agent Pre-migration}

To demonstrate the robustness of our proposed solution under different parameter settings, we test it across various environmental parameters. As shown in Fig. \ref{fig:task}, we validate the effectiveness of our approach by varying the size of AI agent migration tasks. Specifically, we evaluate the total latency of MAPPO and other baseline algorithms under different task sizes, ranging from $25\:\rm{MB}$ to $200\:\rm{MB}$. From Fig. \ref{fig:task}, we can observe that as the size of AI agents increases, the total latency of AI agent migration also increases. However, MAPPO consistently outperforms the baseline algorithms. Specifically, our proposed approach reduces latency by up to $53\%$, $41.9\%$, $38.2\%$, $43.4\%$, $38.5\%$, $43.2\%$, $43.8\%$, and $43.4\%$ across different sizes compared with the baseline methods. Additionally, latency increases with the size of AI agents, which can be attributed to the fact that larger sizes increase the processing latency required by RSUs. Furthermore, when RSUs are under attack, the impact of attacks is amplified by larger sizes of AI agents.

\subsection{Performance on DDoS Defense Schemes}

To demonstrate the effectiveness of the proposed solution when facing different types of DDoS attacks, we conduct a comparison of the average total migration latency under various DDoS attack scenarios. As shown in Fig. \ref{fig:type}, the MAPPO algorithm achieves the lowest migration latency across all three attack scenarios, proving that our solution can effectively mitigate the impact of DDoS attacks. This ensures the secure migration of AI agents while simultaneously reducing latency. To demonstrate the effectiveness of our proposed algorithm under various attack scenarios, we compare it with several baselines under different types of DDoS attacks. As shown in Figs. \ref{dir} - \ref{hyb}, the average test rewards during the convergence process of the proposed MAPPO algorithm consistently surpass those of other baselines. Moreover, defense performance under direct DDoS attacks, indirect DDoS attacks, and hybrid DDoS attacks improved by 15\%, 25\%, and 35\%, respectively, indicating that the proposed method effectively mitigates the negative impact of various DDoS attacks during AI agent migration.

\subsection{Performance on Trust Assessment Mechanism}

We utilize FBR and BR metrics to evaluate the reliability of the proposed trust assessment mechanism. As illustrated in Fig. \ref{fbr}, the declining FBR demonstrates that the proposed trust assessment mechanism continually learns and improves the accuracy in identifying and banning malicious RSUs. Meanwhile, as shown in Fig. \ref{br}, the BR metric steadily increases with the number of iterations, indicating the capability of the trust assessment mechanism to effectively detect and exclude malicious RSUs. This improvement stems from the behavior of malicious RSUs, which frequently transmit tampered or stolen private information to attacking devices, enabling attackers to successfully launch DDoS attacks on target RSUs. This results in a significant increase in malicious scores of RSUs. The banning threshold is adaptively adjusted based on changes in malicious scores and BR, leading to a continuous increase in BR. The numerical results show that the trust assessment mechanism improves over time in identifying malicious RSUs and also demonstrates excellent performance in enhancing overall system reliability and effectively mitigating potential attacks.

\section{Conclusion}
\label{sec:c}
\label{section5}

In this paper, we have investigated the pre-migration strategies of AI agents to defend against potential network attacks during migration. Specifically, we have proposed an online AI agent migration framework in vehicular metaverses, exploring potential DDoS and malicious RSU attacks during the AI agent migration process. To counter DDoS attacks, we have proposed a MAPPO-based online AI agent pre-migration algorithm, designed to achieve optimal pre-migration strategies while minimizing overall migration latency. Furthermore, we have developed a trust assessment mechanism to protect against malicious RSU attacks. Numerical results demonstrate that the proposed solutions can achieve secure AI agent migration with minimal latency. For future work, we will extend our research to address a wider range of network attacks during AI agent migration, aiming to provide more comprehensive and efficient protection against diverse network threats in vehicular metaverses.

\bibliographystyle{IEEEtran}
\bibliography{ref}

\begin{thebibliography}{10}
\providecommand{\url}[1]{#1}
\csname url@samestyle\endcsname
\providecommand{\newblock}{\relax}
\providecommand{\bibinfo}[2]{#2}
\providecommand{\BIBentrySTDinterwordspacing}{\spaceskip=0pt\relax}
\providecommand{\BIBentryALTinterwordstretchfactor}{4}
\providecommand{\BIBentryALTinterwordspacing}{\spaceskip=\fontdimen2\font plus
\BIBentryALTinterwordstretchfactor\fontdimen3\font minus \fontdimen4\font\relax}
\providecommand{\BIBforeignlanguage}[2]{{%
\expandafter\ifx\csname l@#1\endcsname\relax
\typeout{** WARNING: IEEEtran.bst: No hyphenation pattern has been}%
\typeout{** loaded for the language `#1'. Using the pattern for}%
\typeout{** the default language instead.}%
\else
\language=\csname l@#1\endcsname
\fi
#2}}
\providecommand{\BIBdecl}{\relax}
\BIBdecl

\bibitem{9944868}
M.~Xu, W.~C. Ng, W.~Y.~B. Lim, J.~Kang, Z.~Xiong, D.~Niyato, Q.~Yang, X.~Shen, and C.~Miao, ``A {Full} {Dive} {Into} {Realizing} the {Edge-Enabled} {Metaverse}: Visions, {Enabling Technologies, and Challenges},'' \emph{IEEE Communications Surveys \& Tutorials}, vol.~25, no.~1, pp. 656--700, 2023.

\bibitem{10254627}
J.~Kang, J.~Wen, D.~Ye, B.~Lai, T.~Wu, Z.~Xiong, J.~Nie, D.~Niyato, Y.~Zhang, and S.~Xie, ``Blockchain-empowered federated learning for healthcare metaverses: User-centric incentive mechanism with optimal data freshness,'' \emph{IEEE Transactions on Cognitive Communications and Networking}, vol.~10, no.~1, pp. 348--362, 2024.

\bibitem{10401029}
D.~S. Sarwatt, Y.~Lin, J.~Ding, Y.~Sun, and H.~Ning, ``Metaverse for {Intelligent Transportation Systems (ITS)}: {A Comprehensive Review of Technologies, Applications, Implications, Challenges and Future Directions},'' \emph{IEEE Transactions on Intelligent Transportation Systems}, vol.~25, no.~7, pp. 6290--6308, 2024.

\bibitem{zhang2024}
S.~Zhang, X.~Wang, W.~Zhang, Y.~Chen, L.~Gao, D.~Wang, W.~Zhang, X.~Wang, and Y.~Wen, ``{Mutual Theory of Mind in Human-AI Collaboration: An Empirical Study with LLM-driven AI Agents in a Real-time Shared Workspace Task},'' \emph{arXiv preprint arXiv:2409.08811}, 2024.

\bibitem{9973495}
X.~Huang, W.~Zhong, J.~Nie, Q.~Hu, Z.~Xiong, J.~Kang, and T.~Q.~S. Quek, ``{Joint User Association and Resource Pricing for Metaverse: Distributed and Centralized Approaches},'' in \emph{2022 IEEE 19th International Conference on Mobile Ad Hoc and Smart Systems (MASS)}, 2022, pp. 505--513.

\bibitem{9880566}
Y.~Jiang, J.~Kang, D.~Niyato, X.~Ge, Z.~Xiong, C.~Miao, and X.~Shen, ``Reliable {Distributed Computing for Metaverse}: A {Hierarchical Game-Theoretic Approach},'' \emph{IEEE Transactions on Vehicular Technology}, vol.~72, no.~1, pp. 1084--1100, 2023.

\bibitem{YuezhongMetaverse}
Y.~Zhong, J.~Wen, J.~Zhang, J.~Kang, Y.~Jiang, Y.~Zhang, Y.~Cheng, and Y.~Tong, ``Blockchain-assisted twin migration for vehicular metaverses: A game theory approach,'' \emph{Transactions on Emerging Telecommunications Technologies}, vol.~34, no.~12, p. e4856, 2023.

\bibitem{10.1007/978-981-16-8664-1_16}
A.~Gaurav, B.~B. Gupta, and K.~T. Chui, ``{Edge Computing-Based DDoS Attack Detection for Intelligent Transportation Systems},'' in \emph{Cyber Security, Privacy and Networking}, D.~P. Agrawal, N.~Nedjah, B.~B. Gupta, and G.~Martinez~Perez, Eds.\hskip 1em plus 0.5em minus 0.4em\relax Singapore: Springer Nature Singapore, 2022, pp. 175--184.

\bibitem{jsan12040051}
K.~B. Adedeji, A.~M. Abu-Mahfouz, and A.~M. Kurien, ``{DDoS Attack and Detection Methods in Internet-Enabled Networks: Concept, Research Perspectives, and Challenges},'' \emph{Journal of Sensor and Actuator Networks}, vol.~12, no.~4, 2023.

\bibitem{9829332}
T.~Zhang, C.~Xu, P.~Zou, H.~Tian, X.~Kuang, S.~Yang, L.~Zhong, and D.~Niyato, ``{How to Mitigate DDoS Intelligently in SD-IoV: A Moving Target Defense Approach},'' \emph{IEEE Transactions on Industrial Informatics}, vol.~19, no.~1, pp. 1097--1106, 2023.

\bibitem{10018506}
T.~Zhang, C.~Xu, B.~Zhang, X.~Li, X.~Kuang, and L.~A. Grieco, ``{Towards Attack-Resistant Service Function Chain Migration: A Model-Based Adaptive Proximal Policy Optimization Approach},'' \emph{IEEE Transactions on Dependable and Secure Computing}, vol.~20, no.~6, pp. 4913--4927, 2023.

\bibitem{8293801}
Z.~Abdollahi~Biron, S.~Dey, and P.~Pisu, ``{Real-Time Detection and Estimation of Denial of Service Attack in Connected Vehicle Systems},'' \emph{IEEE Transactions on Intelligent Transportation Systems}, vol.~19, no.~12, pp. 3893--3902, 2018.

\bibitem{10.1007/s10586-023-04035-5}
S.~Al-Eidi, O.~Darwish, Y.~Chen, M.~Maabreh, and Y.~Tashtoush, ``{A deep learning approach for detecting covert timing channel attacks using sequential data},'' \emph{Cluster Computing}, vol.~27, no.~2, pp. 1655--1665, jun 2023.

\bibitem{kang2024}
Y.~Kang, J.~Wen, J.~Kang, T.~Zhang, H.~Du, D.~T. Niyato, R.~Yu, and S.~Xie, ``{Hybrid-Generative Diffusion Models for Attack-Oriented Twin Migration in Vehicular Metaverses},'' \emph{ArXiv}, vol. abs/2407.11036, 2024.

\bibitem{10070406}
D.~B. Rawat and H.~El~Alami, ``Metaverse: {Requirements, Architecture, Standards, Status, Challenges, and Perspectives},'' \emph{IEEE Internet of Things Magazine}, vol.~6, no.~1, pp. 14--18, 2023.

\bibitem{9815180}
W.~Y.~B. Lim, Z.~Xiong, D.~Niyato, X.~Cao, C.~Miao, S.~Sun, and Q.~Yang, ``{Realizing the Metaverse with Edge Intelligence: A Match Made in Heaven},'' \emph{IEEE Wireless Communications}, vol.~30, no.~4, pp. 64--71, 2023.

\bibitem{10286997}
S.~Li, X.~Lin, J.~Wu, W.~Zhang, and J.~Li, ``{Digital Twin and Artificial Intelligence-Empowered Panoramic Video Streaming: Reducing Transmission Latency in the Extended Reality-Assisted Vehicular Metaverse},'' \emph{IEEE Vehicular Technology Magazine}, vol.~18, no.~4, pp. 56--65, 2023.

\bibitem{wen2023}
J.~Wen, J.~Kang, Z.~Xiong, Y.~Zhang, H.~Du, Y.~Jiao, and D.~Niyato, ``{Task Freshness-aware Incentive Mechanism for Vehicle Twin Migration in Vehicular Metaverses},'' in \emph{2023 IEEE International Conference on Metaverse Computing, Networking and Applications (MetaCom)}, 2023, pp. 481--487.

\bibitem{10250875}
J.~Feng and J.~Zhao, ``{Resource Allocation for Augmented Reality Empowered Vehicular Edge Metaverse},'' \emph{IEEE Transactions on Communications}, pp. 1--1, 2023.

\bibitem{9491087}
Y.~Lu, S.~Maharjan, and Y.~Zhang, ``{Adaptive Edge Association for Wireless Digital Twin Networks in 6G},'' \emph{IEEE Internet of Things Journal}, vol.~8, no.~22, pp. 16\,219--16\,230, 2021.

\bibitem{kang2024tinymultiagentdrltwins}
J.~Kang, Y.~Zhong, M.~Xu, J.~Nie, J.~Wen, H.~Du, D.~Ye, X.~Huang, D.~Niyato, and S.~Xie, ``{Tiny Multi-Agent DRL for Twins Migration in UAV Metaverses: A Multi-Leader Multi-Follower Stackelberg Game Approach},'' \emph{IEEE Internet of Things Journal}, vol.~PP, pp. 1--1, 06 2024.

\bibitem{BENJABALLAH2020107099}
W.~{Ben Jaballah}, M.~Conti, and C.~Lal, ``{Security and design requirements for software-defined VANETs},'' \emph{Computer Networks}, vol. 169, p. 107099, 2020.

\bibitem{luo2023}
X.~Luo, J.~Wen, J.~Kang, J.~Nie, Z.~Xiong, Y.~Zhang, Z.~Yang, and S.~Xie, ``{Privacy Attacks and Defenses for Digital Twin Migrations in Vehicular Metaverses},'' \emph{IEEE Network}, vol.~37, no.~6, pp. 82--91, 2023.

\bibitem{RePEc}
S.~Sultan, Q.~Javaid, A.~J. Malik, F.~Al-Turjman, and M.~Attique, ``{Collaborative-trust approach toward malicious node detection in vehicular ad hoc networks},'' \emph{Environment, Development and Sustainability: A Multidisciplinary Approach to the Theory and Practice of Sustainable Development}, vol.~24, no.~6, pp. 7532--7550, 2022.

\bibitem{10.1007/978-3-031-19211-1_24}
M.~Li, G.~Zhao, and R.~Lai, ``{A Scalable Blockchain-Based Trust Management Strategy for\&nbsp;Vehicular Networks},'' in \emph{Wireless Algorithms, Systems, and Applications: 17th International Conference, WASA 2022, Dalian, China, November 24-26, 2022, Proceedings, Part III}.\hskip 1em plus 0.5em minus 0.4em\relax Berlin, Heidelberg: Springer-Verlag, 2022, pp. 285--295.

\bibitem{mobility}
S.~Shafi and V.~Ratnam, ``{A Trust Based Energy and Mobility Aware Routing Protocol to Improve Infotainment Services in VANETs},'' \emph{Peer-to-Peer Networking and Applications}, vol.~15, pp. 1--16, 01 2022.

\bibitem{10234402}
T.~Zhang, C.~Xu, Y.~Lian, H.~Tian, J.~Kang, X.~Kuang, and D.~Niyato, ``{When Moving Target Defense Meets Attack Prediction in Digital Twins: A Convolutional and Hierarchical Reinforcement Learning Approach},'' \emph{IEEE Journal on Selected Areas in Communications}, vol.~41, no.~10, pp. 3293--3305, 2023.

\bibitem{10534545}
F.~Reza, ``{DDoS-Net: Classifying DDoS Attacks in Wireless Sensor Networks with Hybrid Deep Learning},'' in \emph{2024 6th International Conference on Electrical Engineering and Information \& Communication Technology (ICEEICT)}, 2024, pp. 487--492.

\bibitem{Obaidat2020}
M.~Obaidat, M.~Khodjaeva, J.~Holst, and M.~Ben~Zid, \emph{{{Security and Privacy Challenges in Vehicular Ad Hoc Networks}}}.\hskip 1em plus 0.5em minus 0.4em\relax Cham: Springer International Publishing, 2020, pp. 223--251.

\bibitem{9152970}
A.~Masood, D.~S. Lakew, and S.~Cho, ``{Security and Privacy Challenges in Connected Vehicular Cloud Computing},'' \emph{IEEE Communications Surveys \& Tutorials}, vol.~22, no.~4, pp. 2725--2764, 2020.

\bibitem{8539991}
X.~Wang, Z.~Ning, M.~Zhou, X.~Hu, L.~Wang, Y.~Zhang, F.~R. Yu, and B.~Hu, ``{Privacy-Preserving Content Dissemination for Vehicular Social Networks: Challenges and Solutions},'' \emph{IEEE Communications Surveys \& Tutorials}, vol.~21, no.~2, pp. 1314--1345, 2019.

\bibitem{10.1007/978-981-99-3481-2_37}
M.~F. Singha and R.~Patgiri, \emph{{A Survey on DDoS Detection Using Deep Learning in Software Defined Networking}}.\hskip 1em plus 0.5em minus 0.4em\relax Singapore: Springer Nature Singapore, 11 2023, pp. 479--494.

\bibitem{9488287}
N.~V. Abhishek, M.~N. Aman, T.~J. Lim, and B.~Sikdar, ``{DRiVe: Detecting Malicious Roadside Units in the Internet of Vehicles With Low Latency Data Integrity},'' \emph{IEEE Internet of Things Journal}, vol.~9, no.~5, pp. 3270--3281, 2022.

\bibitem{10.114}
C.~E. Shannon, ``{A mathematical theory of communication},'' \emph{SIGMOBILE Mob. Comput. Commun. Rev.}, vol.~5, no.~1, pp. 3--53, Jan. 2001.

\bibitem{10415630}
J.~Kang, J.~Chen, M.~Xu, Z.~Xiong, Y.~Jiao, L.~Han, D.~Niyato, Y.~Tong, and S.~Xie, ``{UAV-Assisted Dynamic Avatar Task Migration for Vehicular Metaverse Services: A Multi-Agent Deep Reinforcement Learning Approach},'' \emph{IEEE/CAA Journal of Automatica Sinica}, vol.~11, no.~2, pp. 430--445, 2024.

\bibitem{10015746}
A.~Boualouache and T.~Engel, ``{A Survey on Machine Learning-Based Misbehavior Detection Systems for 5G and Beyond Vehicular Networks},'' \emph{IEEE Communications Surveys \& Tutorials}, vol.~25, no.~2, pp. 1128--1172, 2023.

\bibitem{PISINGER20052271}
X.~Sun and N.~Ansari, ``Primal: Profit maximization avatar placement for mobile edge computing,'' in \emph{2016 IEEE International Conference on Communications (ICC)}, 2016, pp. 1--6.

\bibitem{dewitt2020independentlearningneedstarcraft}
C.~S.~D. Witt, T.~Gupta, D.~Makoviichuk, V.~Makoviychuk, P.~H.~S. Torr, M.~Sun, and S.~Whiteson, ``Is independent learning all you need in the starcraft multi-agent challenge?'' \emph{ArXiv}, vol. abs/2011.09533, 2020.

\bibitem{siddiqua}
F.~Siddiqua and M.~Jahan, ``{A Trust-Based Malicious RSU Detection Mechanism in Edge-Enabled Vehicular Ad Hoc Networks},'' \emph{ArXiv}, vol. abs/2407.11036, 2022.

\bibitem{KERRACHE20181}
C.~A. Kerrache, A.~Lakas, N.~Lagraa, and E.~Barka, ``{UAV-assisted technique for the detection of malicious and selfish nodes in VANETs},'' \emph{Vehicular Communications}, vol.~11, pp. 1--11, 2018.

\end{thebibliography}


\end{document}